\documentclass[journal]{IEEEtran}
\usepackage{graphicx}              
\usepackage{amsmath,amssymb}
\usepackage{mathtools,cuted}
\usepackage{breqn}
\usepackage{lipsum}
\usepackage{cite}
\usepackage{multicol}
\usepackage{xcolor}

\begin{document}

\title{\LARGE LSTM-Based Distributed Conditional Generative Adversarial Network For Data-Driven 5G-Enabled Maritime UAV Communications}
\author{Iftikhar Rasheed, Muhammad Asif, Asim Ihsan, Wali Ullah Khan, Manzoor Ahmed and Khaled Rabie\thanks{Iftikhar Rasheed is with the Department of Information and Communication Engineering, The Islamia University of Bahawalpur, Pakistan (email: iftikhar.rasheed@iub.edu.pk).

Muhammad Asif is with Guangdong Key Laboratory of Intelligent Information Processing, College of Electronics and Information Engineering, Shenzhen Univerisyt, Shenzhen, Guangdong, China (email: masif@szu.edu.cn).

Asim Ihsan is with the Department of Information and Communication Engineering, Shanghai Jiao Tong University, Shanghai China (email: ihsanasim@sjtu.edu.cn).

Wali Ullah Khan is with Interdisciplinary Centre for Security, Reliability and Trust (SnT), University of Luxembourg, 1855 Luxembourg City, Luxembourg (email: waliullah.khan@uni.lu).

Manzoor Ahmed is with the College of Computer Science and Technology, Qingdao University, Qingdao 266071, China (email: manzoor.achakzai@gmail.com).

 Khaled Rabie is with the Department of Engineering, Manchester Metropolitan University, UK and with the Department of Electrical and Electronic Engineering, University of Johannesburg, P.O. Box 17011 Doornfontein, Johannesburg 2028, South Africa (email: k.rabie@mmu.ac.uk).
}
}

\markboth{IEEE Transactions on Intelligent Transportation Systems, vol. YY, no. ZZ, 2022}%
{Shell \MakeLowercase{\textit{et al.}}: Bare Demo of IEEEtran.cls for IEEE Journals}

\maketitle

\begin{abstract} 
5G enabled maritime unmanned aerial vehicle (UAV) communication is one of the important applications of 5G wireless network which requires minimum latency and higher reliability to support mission-critical applications. Therefore, lossless reliable communication with a high data rate is the key requirement in modern wireless communication systems. These all factors highly depend upon channel conditions. In this work, a channel model is proposed for air-to-surface link exploiting millimeter wave (mmWave) for 5G enabled maritime unmanned aerial vehicle (UAV) communication. Firstly, we will present the formulated channel estimation method which directly aims to adopt channel state information (CSI) of mmWave from the channel model inculcated by UAV operating within the Long Short Term Memory (LSTM)-Distributed Conditional generative adversarial network (DCGAN) i.e. (LSTM-DCGAN) for each beamforming direction. Secondly, to enhance the applications for the proposed trained channel model for the spatial domain, we have designed an LSTM-DCGAN based UAV network, where each one will learn mmWave CSI for all the distributions. Lastly, we have categorized the most favorable LSTM-DCGAN training method and emanated certain conditions for our UAV network to increase the channel model learning rate. Simulation results have shown that the proposed LSTM-DCGAN based network is vigorous to the error generated through local training. A detailed comparison has been done with the other available state-of-the-art CGAN network architectures i.e. stand-alone CGAN (without CSI sharing), Simple CGAN (with CSI sharing), multi-discriminator CGAN, federated learning CGAN and DCGAN. Simulation results have shown that the proposed LSTM-DCGAN structure demonstrates higher accuracy during the learning process and attained more data rate for downlink transmission as compared to the previous state of artworks. 
\end{abstract}
\begin{IEEEkeywords}
Maritime Unmanned aerial vehicle (UAV) Communication, 5G, LSTM, mmWave, DCGAN.
\end{IEEEkeywords}

\section{Introduction}
 Advanced cellular technologies are intended to enhance wireless communication in several aspects such as coverage extension, massive connectivity, transmission latency, spectral efficiency, reliability, and energy efficiency \cite{ihsan2022energy,9739970,hasan2022securing}. For 5G and beyond wireless systems, the millimeter wave (mmWave) frequency spectrum is vital as it will enable various next-generation wireless system applications. These applications are airborne wireless networks, vehicular communication for autonomous driving, and maritime unmanned aerial vehicles (UAVs), which intend to provide higher data rates with much lower network latency \cite{jijo2021comprehensive},\cite{rasheed2021intelligent}, \cite{zheng2022graph}, \cite{pan2022artificial}. However, the main issue with mmWave based communication is higher attenuation loss. To overcome this problem, multiple-input multiple-output (MIMO) with highly-directional beamforming is used to increase the cell throughput with reliability. Moreover, mmWave has a shorter coherence time than the sub-6 GHz communication. Hence, 5G-mmWave communication links are more time-sensitive, and they need frequent channel measurements. The highly-directional beamforming requires the exact information angle of arrival (AoAs) and angle of departure (AoDs) to establish perfect beam alignment between transceivers. This channel estimation and beam training process can increase communication overhead and impact spectrum usage. Therefore, for better network performance, it is very important to formulate an effective model with its MIMO channels for the 5G-mmWave link \cite{noh2020fast}, \cite{abbasi2022lidar}, \cite{babbar2022lbsmt}.

Incorporating Maritime UAVs with 5G-mmWave can increase data rates in both uplink and downlink transmission. However, UAV based network over 5G-mmWave is a challenging job due to variable channel conditions during altitude mobility \cite{dabiri2020analytical}. In \cite{zhang2019reflections} UAV based base-station (BS) was suggested, which can easily adjust its location using the of results generated through entailed deep neural networks. One of the major differences between terrestrial channel estimation and air to ground channel estimation (A2G) is that terrestrial channel modeling is relatively easy with a known condition, whereas, in the A2G scenario, channel condition varies with time, altitude, weather, speed, and geography. Therefore, the channel model obtained from UAV at a given condition and time cannot be generalized to other environments \cite{yang2019generative}. Due to this, conventional channel estimation techniques like ray tracing, geographically estimated statistical models are not enough in A2G channel modeling for 5G connected UAVs. 

Furthermore, mostly available A2G channel models are generally rectified at the sub-6 GHz range, whereas, standard A2G models for 5G-mmWave and their respective data are not sufficient \cite{rasheed2020intelligent}. Therefore, for efficient and lossless transmission with the maximized data rate, there is a need for a closely related A2G channel model. In this A2G based scenario, a data-driven scheme can be employed where multiple UAVs operating in non-overlapping regions can train their respective model by extracting CSI during cellular service from their regions to formulate a whole channel model (based on altitude, speed, geography, etc.) for the entire environment (spatial-temporal map) each UAV will share their channel model parameters with other UAVs for better learning and estimation of A2G mmWave link \cite{khan2022opportunities}. 

\subsection{Related Work}

5G-mmWave based communication offers higher bandwidth with much lower network latency. Previously the wireless communication spectrum was aimed at the utilizing the spectrum between 1.9 GHz to 2.2 GHz. However, the availability of spectrum for higher data rate and IoT/IoV applications in the future wireless communication standard is one of the key challenges. Ubiquitous channel models were only estimated for the existing band, but literature targeting future applications aims in exploiting the mmWave band. Therefore, using a data-driven method for the mmWave band, numerous statistical models were proposed in \cite{akdeniz2014millimeter},\cite{dong2019deep},\cite{polese2020experimental} and \cite{cheng2020modeling} etc. Unlike the data-driven approach, conventional methods of spatial-temporal correlation and compressed sensing were explored for describing MIMO-based directional mmWave communication \cite{akdeniz2014millimeter}, \cite{asif2019construction}. 

\begin{table}[!t]\footnotesize
\renewcommand{\arraystretch}{1.1}
\centering
\caption{Previous Work Limitations}
\begin{tabular}{|p{0.8in}|p{2.2in}|} \hline
\textbf{Previous Works} & \textbf{Drawbacks} \\ \hline
Cheng et al. \cite{cheng2020modeling}
 & \begin{itemize}
     \item A statistical conventional method. 
     \item Only used for limited applications with very poor latency. 
 \end{itemize}
\\ \hline
Zhang et al. \cite{zhang2021distributed}
 & \begin{itemize}
 \item Uses DCGAN Approach 
 \item But it had higher convergence and latency time.
 \end{itemize} \\ \hline
 Ye et al. \cite{ye2020deep} & \begin{itemize}
     \item MD-CGAN distributed learning scheme.
     \item Very high latency and poor convergence. 
 \end{itemize}
  \\ \hline
Xia et al. \cite{xia2020generative} & \begin{itemize} \item Independent CGAN at each UAV without cooperation. \item Offers poor latency and network stability. \end{itemize} \\ \hline
Elbir et al. \cite{elbir2021federated} & \begin{itemize}
    \item An FLCGAN method with high latency.
    \item Less stable link formation for 5G UAVs.
\end{itemize} \\ \hline
\end{tabular}
\end{table}

Another proposal in \cite{dong2019deep} utilized spatial frequency neural networks (SF-CNN) for approximating MIMO-based communication under mmWave frequency range. Whereas, \cite{alkhateeb2019deepmimo} shared dataset parameters of transmission link held via mmWave channel using deep learning scheme. Moreover, in all of these works their results were obtained using mmWave under terrestrial environment, we can say that these studies only meant for channels just above the surface. 

However, the scenario is quite different among terrestrial communication and A2G communication links. It differs in wave propagation, Doppler effect, fading, and multipath. Therefore, some other schemes with different methods were proposed to estimate the mmWave channel model in airborne transmission scenarios \cite{yang2019generative},\cite{cheng2020modeling} and \cite{han2016two}. Considering the UAV scenario, \cite{khawaja2019survey} comprehensively reviewed previously proposed A2G channel models and suggested the pros and cons of each with promising future research direction. Authors in \cite{yang2019generative} utilized neural networks for microwave and mmWave band to characterize channel model with UAV in A2G communication through data set training obtained by a generative neural network. 

In \cite{cheng2020modeling} presented a path loss model based on signal propagation between air to air scenarios using UAV at 60 GHz. However, \cite{han2016two} followed the same traditional ray-tracing method at 28 GHz to estimate a geometrically stochastic model for the A2G scenario using UAV. The wave propagation model and some of its parameters (received signal strength, root mean square delay, and multipath) were deeply analyzed in \cite{polese2020experimental} using simulation at 28 GHz and 60 GHz. Secondly, in this approach they also developed a channel model using USRP to characterize A2G based mmWave channel for inter UAV collaboration. 

All these mentioned previous state of art works  have focused on estimating mmWave channel model for limited environment and supports very few applications. Consequently, these available models cannot be generalized for all applications of A2G communication using mmWave in surrounding environment. 

By exploring a wide area of application for channel modeling in airborne or A2G communication, jointly modeling dispersed channels together is one of the core areas of research. Unlike other schemes which require nodes to share CSI with the base station (BS) obtained by pilot signal, \cite{khawaja2017uav} proposed federated learning (FL) approach trained through a convolutional neural network (CNN) using a dataset extracted from CSI by sending a pilot signal. However, the FL approach is centralized in nature and requires a controller to manage the overall network. But airborne A2G case is fully distributed in nature, therefore we cannot imply \cite{khawaja2017uav} in our modeling. Another work in \cite{elbir2020federated} considered distributed networks and employed machine learning (ML) in time-variant channels by exchanging information continuously.

Consequently, this continuous data sharing brings heavy costs in communication in terms of bandwidth utilization. Unlike previous work \cite{khawaja2017uav} utilized FL based learning and SF-CNN, the proposal in \cite{park2020communication} focused on a generative learning-based model for the mmWave spectrum channel. Therefore, in traditional channel models like ray tracing and other statistical models, core channel knowledge with wave propagation information is necessary. Hence, the model in \cite{zhang2019reflections} hereby proposed an adversarial generative network (GAN) which autonomously support estimating a channel model. 

If we consider \cite{park2020communication}, it solely focuses on introducing GAN, where deep neural network process acquired raw data from the transmitted pilot signal and autonomously configure its encoder, puncture rate, interleaver, and modulator, depending upon CSI obtained from the received pilot signal. From the above discussion, we can conclude that all these previous proposals focused on autoconfiguration of baseband signal processing block depending on obtained CSI and some were only targeting terrestrial network architecture in centralized manner, which is quite out of domain in our research of estimating A2G communication under mmWave spectrum. 

Therefore, we can infer that there is still deficiency of fully distributed generative learning model to overcome data driven based channel modeling. We in this brief will explain fully decentralized generative learning model to describe A2G mmWave communication. Table I shows limitations of the previous works that are needed to be addressed.

The major focus of this proposed work is providing an infrastructure that uses 5G-mmWave based UAV network incorporating intelligent technique i.e. LSTM-DCGAN. Its a data driven approach coupled with machine learning method. The key consideration is UAV-based fully distributed network communicating in the air to the ground scenario using mmWave spectrum for 5G enabled maritime UAV network. The proposed work will also focus by estimating its data-driven channel modeling. Particularly, we will also exploit machine learning to train a model using LSTM-DCGAN, where UAVs will learn cooperatively from each other’s dataset. 

\subsection{Major Contribution of the Proposed Work}
The main contributions for the proposed work are:
\begin{itemize}

\item First of all a CSI extraction technique is presented to acquire the real-time CSI of 5G spectrum in the A2G scenario, where each 5G enabled UAV will train the respective model at beamforming direction using LSTM-DCGAN. In previous works \cite{cheng2020modeling},\cite{zhang2021distributed},\cite{ye2020deep},\cite{xia2020generative},\cite{elbir2021federated}, all of these methods lacks systematic and effective approach for CSI extraction. Thus, these methods did not produce an effective CSI for 5G UAV networks needed for mission critical applications especially for maritime UAV communication. 

\item Secondly, this work improves the spectrum learning by effectively considering spatial-temporal domain. This is incorporated with cooperative learning model developed using LSTM-DCGAN approach. This decentralized learning approach facilitate each airborne device to train itself from the distributed dataset. Moreover, to avoid sharing the actual estimated channel model, we facilitated each 5G enabled maritime UAV to share only mock information obtained from the channel. Hence, this approach is fully decentralized and does not require any server to handle its operations. Whereas, previous approaches were not decentralized. Only \cite{zhang2021distributed} presented an effective method using DCGAN but it had higher convergence and latency time.

\item Effective learning is achieved through logically formulated the intersection probability of decentralized LSTM-DCGAN learning. Subsequently, we further derived necessary boundary conditions to facilitate ideal 5G enabled maritime  UAV to UAV link, which enhances learning rate within cooperative spectrum modeling. Previous methods were not able to meet effective learning as number of 5G enabled maritime UAV increases. \cite{zhang2021distributed} although showed to achieve a constant average data rate with increase in number of connected maritime UAVs but average data rate achieved was quite low with high latency.

\item For better performance evaluation of the proposed work, we have simulated our distributed network and applied LSTM-DCGAN based learning for airborne 5G enabled maritime UAVs. The result shows that proposed idea of a fully distributed network with the LSTM-DCGAN model is more vigorous and shows less error during the model simulation. For comparison, we have opted for the available CGAN model deprived of sharing dataset samples \cite{xia2020generative} and are non-distributed in design like multi-discriminator CGAN \cite{cheng2020modeling}, FL-based CGAN methods \cite{elbir2021federated}, MD-CGAN distributed learning scheme \cite{ye2020deep}, and Zhang et al. \cite{zhang2021distributed}. 

\item Simulations have shown that the proposed LSTM-DCGAN based model for 5G enabled maritime UAVs is a more accurate and reliable channel model with high data rate and lower latency. Moreover, it has better convergence and able to handle high number of maritime UAVs in an effective manner when compared to the previous state of art works.

\end{itemize}

\begin{figure*}[!t]
\centering
\includegraphics [width=0.80\textwidth]{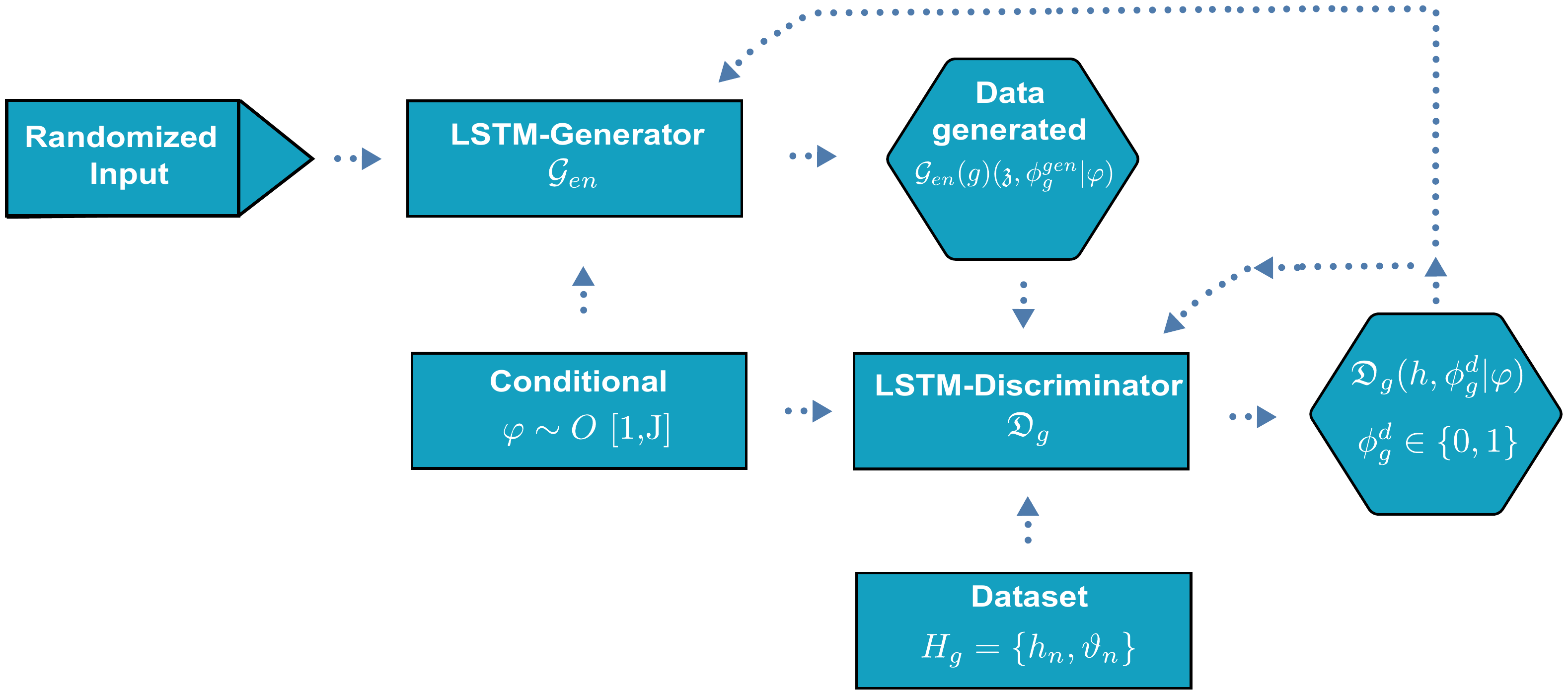}
\caption{System Model Based on Long Short Term Memory (LSTM)-Distributed Conditional generative adversarial network (DCGAN).}
\label{fig:Fig1}
\end{figure*}

\section{5G mmWave Based Communication Model}

Within our communication network prototype, we assumed $\textrm{g}$\textit{ }as a set of UAVs communicating over a downlink to the ground station, we here denote ground station as \textit{GS}. For A2G transmission both ends are communicating with antennas, we made another assumption that UAVs are transmitting through linearly attached array antennas denoted as \textit{L, }whereas UAV's transmitter vector is given by ${\beta }_q\left({\vartheta }^q\right)={\ \left[1,\ e^{i\frac{\pi }{\lambda }{\mathrm{sin} {\vartheta }^q\ }},\ \dots \dots ,\ e^{i\left(L-1\right)\frac{\pi }{\lambda }{\mathrm{sin} {\vartheta }^q\ }}\right]}^Q$, we denote lambda $\lambda $ for the wavelength of carrier frequency and ${\vartheta }^q$is the angle of departure in degrees from $0\sim 2\pi .$ For course with transmitter array, there must be an arranged array of receiving antennas placed linearly represented as \textit{K. }Hence, the receiver in the vector can be mathematically written as ${\beta }_p\left({\vartheta }^p\right)={\ \left[1,\ e^{i\frac{\pi }{\lambda }{\mathrm{sin} {\vartheta }^p\ }},\ \dots \dots ,\ e^{i\left(K-1\right)\frac{\pi }{\lambda }{\mathrm{sin} {\vartheta }^p\ }}\right]}^Q,$ here we denote AoA as ${\vartheta }^p$. We have considered MIMO communication and represents channel matrix as $J\epsilon \mathbb{C}$ with dimension \textit{LxK}. Therefore, now we mathematically denote $J$ matrix as:
\begin{equation} \label{1} 
J=\sum^W_{w=1}{A_{w.}}{\beta }_p\left({\vartheta }^p_w\right)+{\beta }^J_q\left({\vartheta }^q_w\right) 
\end{equation} 

Here we took ${(.)}^J$ as conjugate transpose, we further represented different paths by \textit{W. }in equation 1 we say that $A_w$which belongs to complex channel gain $\mathbb{C}$ i. Whereas, ${\beta }_q$ and ${\beta }_p$ and angles from distinct path \textit{w. }Provided that our A2G communication through the mmWave channel needs a clear line of sight and very sensitive to the path hurdles therefore also have some refraction paths due to which \textit{W} will be quite smaller than \textit{LxK} size. The main benefit rendered through MIMO is the beamforming technique, where a narrow beam of mmWave signal carrying information is transmitted from one end to the other without any fear of receiving delayed multipath caused by refractions and reflections \cite{khawaja2017uav}.

Therefore, to ideally have only one main lobe at both transmitters and receivers instead of sides lobes in the frequency domain, in our A2G communication scenario we supposed that each UAV has a unique directional link to the ground station via beam. Hence, \cite{polese2020experimental} and \cite{swindlehurst2014millimeter} performed experiments at 60GHz by assuming mmWave as beam and suggested that beamforming over mmWave is similar to have only one channel with \textit{W=1 }as a case in a line of light (LOS), reflected none line of sight (NLOS) or complete outage.  \cite{khawaja2018temporal} also support utilizing a single mmWave channel by designing a unique transceiver. Whereas, work in \cite{khalili2020optimal},\cite{khojastepour2020multi} and \cite{sadhu20177} solely assumed to exploit unique and directional beamformed channels for mmWave communications. Now let us assume UAV position as \textit{u} in 3D plane and coordinates of GS as \textit{y. }Then at any time instant \textit{t,} MIMO channel matrix in \eqref{1} for A2G communication can be described as:
\begin{equation} \label{2} 
J\left(u,v,\ t,\ {\vartheta }^q,\ {\vartheta }^p\right)=A\left(u,v,t,\ {\vartheta }^q,\ {\vartheta }^p\right){\beta }_p\left({\vartheta }^p\right)+{\beta }_q({\vartheta }^q) 
\end{equation} 

$A$\textit{ }the channel gain in \eqref{1} and \eqref{2} is determined by simultaneously solving AoA, AoD vectors of mmWave path, and channel spectrum (\textit{u,v,t}). 

\subsection{5G mmWave Channel Estimation}
Sending a pilot signal from the transmitter is a common technique for estimating channel conditions \cite{le2021deep}. However, on receiver received pilot signal describes the CSI in the form of the dataset, which then multiplied with a received symbol to overcome the channel effect. Similarly, in our UAV based A2G mmWave channel, each airborne object transmits a pilot signal with power denoted as $P_w.$ Moreover, as we are assuming beam spectrum for communication, therefore, it is essential to synchronize both UAV and GS over a pre-computed codebook. Now let us define \textit{I }as a codebook length and $(e_i,f_i)$ as ith pair representing beamforming and combing vector in the codebook. At the ground station (GS), for ith training, the received pilot symbol can be given as in:
\begin{equation} \label{3} 
P_{Sym}=\sqrt{{P_w}f^J_i / J_ie_i}+f^J_iN,   
\end{equation} 
We say that \textit{N }is defined as a vector having noise information of the A2G link. Likewise in \cite{khawaja2018temporal} we also assumed that both GS and UAV have substantial information on the radiation pattern of their antenna arrays to compute beamforming ($e_i$), combining vector ($f_i$) along with AoA (${\vartheta }^p_i(f_i)$) and AoD (${\vartheta }^q_i(e_i)$). Now, for obtaining received pilot symbol, we define $\otimes $ as Kronecker product and \textit{Z}(.) be the vectors inside matrix.

\begin{align} \label{4} 
 P_{Sym} = & \sqrt{{P_w}\left(e^Q_i\bigotimes f^J_i\right)Z\left(J_i\right)}+f^J_iN \nonumber\\
 = &\sqrt{{P_w}\left(e^Q_i\bigotimes f^J_i\right)[{\beta }^*_q\left(e_i\right)\bigotimes {\beta }_p\left(f_i\right)]A_i(u,v,t,{\vartheta }_i)}\nonumber\\
& +f^J_iN 
\end{align} 

Afterward, on getting the pilot symbol $P_{Sym}$, the receiver will extract CSI and share this dataset with UAVs through sub-6 uplink \cite{umit2017real}. On the other side base station have prior knowledge of pair vectors $(e_i,f_i)$ from which transmitter and receiver vectors can also be computed. In general, \eqref{4} say that channel gain of the airborne object at coordinate \textit{u }communicating with GS at location \textit{v }at time instant \textit{t }having AoA-AoD pair ${\vartheta }_i$can be obtained through \eqref{5}, whereas estimated error is described by $\widetilde{N_i}$.
\begin{equation} \label{5} 
\widetilde{A_i}\left(u,v,t,{\vartheta }_i\right)=P_{Sym}{\beta}{_i}=A_i\left(u,v,t,{\vartheta }_i\right)+\widetilde{N_i} 
\end{equation} 

In A2G communication, by utilizing \textit{I }distinct antenna pairs in UAVs operating in all orientation can extract CSI and channel gain $\widetilde{A_k}$ in the spatial-temporal domain. Thus, we denote spectrum information of each 5G enabled UAV denoted by $'g'$\textit{ }as a set $H_g=\left\{\ h_n,\ {\vartheta }_n\right\}$ where $n$ is the size of dataset estimated by $gth$ 5G enabled UAV and equivalently we can write channel dataset as $\{u_n,v_n,\ t_n,\ \widetilde{A_n,}{\vartheta }_n\}$ and we already knew that AoA-AoD for each $h_n$ is denoted as ${\vartheta }_n$ in degrees from $0\sim2\pi$.  Also, we represent cumulative dataset acquired from all channels as {\textbar}$H_g|$. Each 5G enabled UAV develop its own estimated model trained from learning CSI dataset $H_g$ acquired from mmWave communication in A2G link.

Conventional channel model estimation techniques have some considerable limitations in the current scenario where spectrum utilization is one of the key challenges. Therefore, regression, ray tracing, and other geographically statistical models are not sufficient in modern applications of mmWave link. One major key factor is that channels are specifically modeled for terrestrial communication considering communication loss incurred by multipath, fading, and interference \cite{rasheed2021intelligent}, \cite{asif2021reduced}, \cite{asif2020jointly}.

However, exploiting the spectrum of 5G-mmWave in A2G link is a new idea for extending its applications in future wireless communication standards \cite{rasheed2020machine}. Also, in the literature there are very limited studies are proposed for estimation of A2G link implying data-driven approach \cite{zhang2021distributed} and \cite{swindlehurst2014millimeter}, specifically focusing on temporal and spatial characteristics of 5G-mmWave spectrum \cite{han2016two}. Considering spatial-temporal domain in MIMO-based channel response, the CSI dataset for training acquired by each airborne UAV is insufficient for the precise approximation of spectral analysis including phase, amplitude, and directional features. In the given case, where the height of the flying object estimates the probability of line of sight from UAV to GS, moreover, the accuracy of approximated 5G-mmWave channel is directly proportional to the size of the channel samples dataset, included in the learning stage.

\begin{figure*}[!t]
\centering
\includegraphics [width=0.80\textwidth]{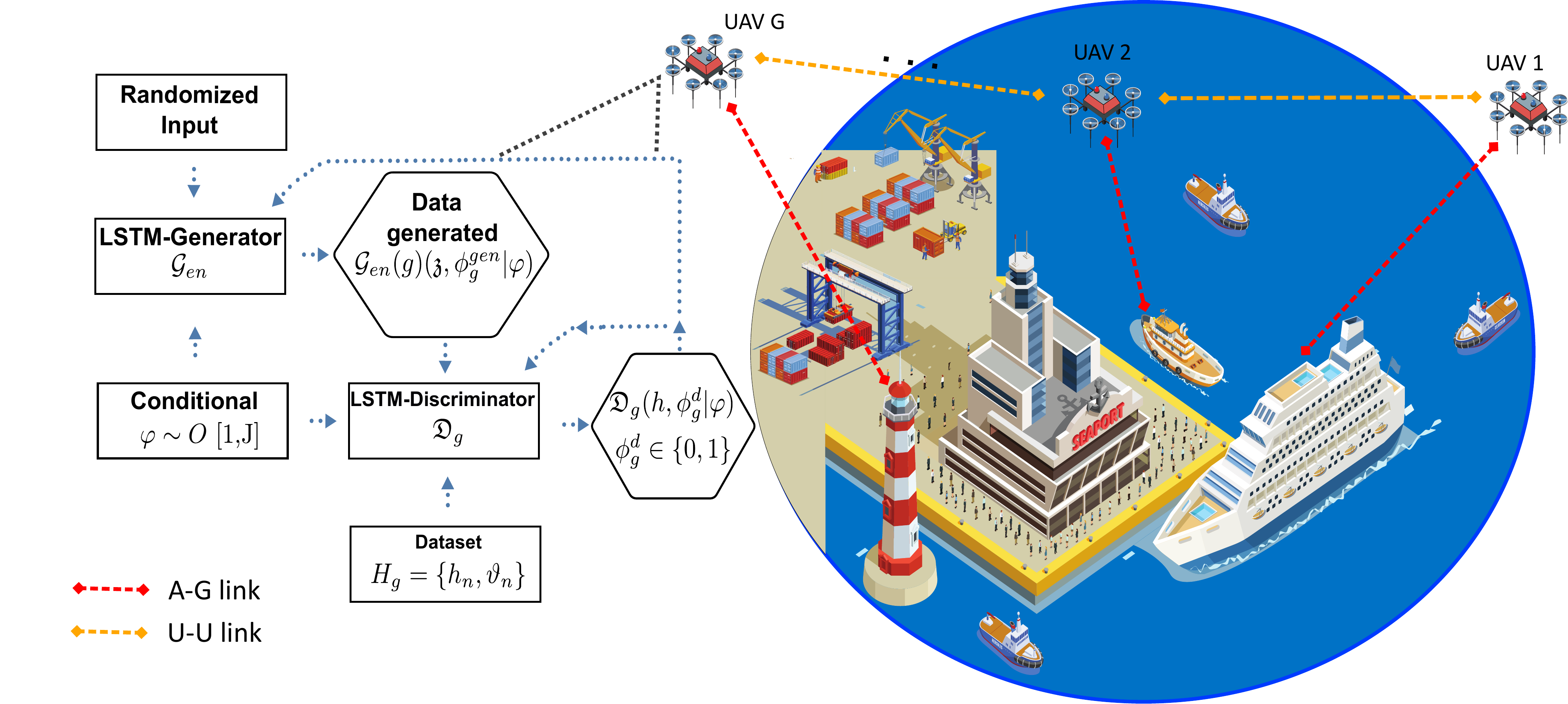}
\caption{Long Short Term Memory-Distributed Conditional generative adversarial network (LSTM-DCGAN) with $G$ number of 5G Enabled Maritime UAVs}
\label{fig:Fig2}
\end{figure*}
Furthermore, we know that 5G link features Vary with respect to time and geographical conditions. Therefore, to develop a generic model of 5G spectrum in the air to ground applications, a large amount of dataset from various perspectives are required for example weather effects throughout the year, number of cellular nodes operating in the region, the altitude of sensing node and number of airborne sensors collecting dataset. These factors are some of the key elements. Hence, the 5G-mmWave channel modeling for A2G application requires an effective mechanism, therefore in this work we will present a deep learning-based data-driven channel model with collaborative CSI sharing among 5G maritime UAVs, operating in large scale spatial and temporal domain.

\section{Distributed 5G Enabled Maritime UAV Channel Model Using LSTM-DCGAN}

\subsection{LSTM-DCGAN for 5G mmWave Channel Modeling}

For \textit{g} number of airborne 5G enabled UAVs extracting CSI in A2G link, there exist \textit{g} learned models trained by each UAV through \textit{g}collected datasets using LSTM networks. Some distinct models are designed to input spatial-temporal pairs during dataset training and generate complex channel gain as in \cite{noh2020fast}, in these models, UAVs can predict any new random inputs if applied to the 5G-mmWave channel. However, from the channel dataset $H_g$, discriminative models will not adopt any additional information except channel gain. From collected data ${(u_n,v_n,\ t_n)}_{\forall n}\subset H_g$, $u,v$\textit{ }coordinates of UAV can be used to point exact location and movement(from initial and last point), whereas, from time \textit{t }we can also estimate the speed factor of this movement.

With the distributed and decentralized structure in the spatial-temporal domain, pairs in a set $H_g$ can facilitate other applications of the trained spectrum model. Hence, on getting motivation from the spatial-temporal domain learning model, we followed to propose a generative method for 5G-mmWave A2G channel estimation.

For UAV communicating with GS at downlink, we jointly calculated the pairwise sequence of AoA-AoD through a pre-computed codebook. We further defined $\vartheta $ (independent of CSI) as prior information of both transmitting and receiving antenna. Afterward, modeling channel for all arriving and departing orientation, we exploited conditional generative adversarial network model as in \cite{feng2018spectrum}, whose mathematical model defines a condition sampler denoted by \textit{O, } whereas generator is represented as ${\mathcal{G}}_{en}$ and ${\mathfrak{O}}_g$ for discriminator. In general, within each training session, condition sampler collect pair of AoA and AoD from available $\mathcal{I}$ uniformly distributed directions $\varphi \sim O[1,\ \mathcal{I}]$, which are also similar for each airborne object. 

Likewise, the LSTM based generator ${\mathcal{G}}_{en\ \left(g\right)}(\mathfrak{z},{\emptyset }^{g_{en}}_g|\varphi )$ assigns random input $\mathfrak{z}$ to channel dataset \textit{H } with condition $\varphi $ and discriminator ${\mathfrak{D}}_g(h,\ {\emptyset }^{d_{\ }}_g|\varphi $) which uses a parameter vector ${\emptyset }^{d_{\ }}_g$ having input channel sample \textit{h } and condition $\varphi $ and generates an output between 0 and 1. As result, if discriminator output is near to 1 then we infer that input sample \textit{h}$=(u,v,t,A)$\textit{ } is likely to be a real dataset whose channel gain is observed when airborne object has position on point \textit{u } and GS location is on \textit{v} at timestamp \textit{t}. 

However, in another case when the discriminator is zero, we denote that the channel sample is fake. This is the reason that generators in airborne objects target to generate channel samples closely related to real data, on the other hand, a discriminator aims to identify real channel samples from the false sample. 

We suppose here $b_g$ be the distribution of the sample in the channel dataset $H_g$ and learned distribution computed by generator function for airborne object \textit{g} is represented by $b^{{\mathcal{G}}_{en}}_g$, whereas the distribution of random input is denoted by $b^{\mathfrak{z}}_g$. Afterward, generator ${\mathcal{G}}_{gen}$ is trained by autonomous LSTM-DCGAN model to elicit channel distribution $b_g$ for each condition defined by $\varphi $. Further, to enumerate the learning accuracy of ${\mathcal{G}}_{gen}$ trained by each UAV \textit{g, }we used ${\mathfrak{D}}_g$ discriminator. 

Therefore, for each UAV \textit{g, }we develop a relation among generator and discriminator for information sharing using the zero-sum method with value function as it is in \cite{feng2018spectrum} and \cite{zhang2021distributed}. The main difference here is that we have LSTM-DCGAN incorporated, whereas these previous methods uses simple deep neural network structure. 

\begin{equation} \label{6}
\begin{split}
R_{g}(\mathfrak{D}_{\mathfrak{g}},{\mathcal{G}}_{en\ (g)})=\frac{1}{\mathcal{I}}\sum_{\mathcal{I}=1}^{\mathcal{I}}\varepsilon_{h\sim b_{g}^{\mathfrak {z}}}[\log{\mathfrak{D}_{g}}(h\mid\varphi_{\mathcal{I}})]-\\ \varepsilon_{{\mathfrak{z}} \sim b_{g}^{\mathfrak{z}}}[\log(1-\mathfrak{D}_{\mathfrak{g}}({\mathcal{G}}_{en\ (g)}({\mathfrak{z}}\mid\varphi_{\mathcal{I}})
\end{split}
\end{equation} 

In the first half portion of the above formula, there will be $\mathcal{I}$ conditions for directions, and real data input with each value of ${\varphi }_{\mathcal{I}}$ \eqref{6} enables the discriminator to generate one at the output. Whereas, in rest of the part $G_{en(g)}$ have a dominant effect on refining the dataset produced by the generator. Overall, two entities ${\mathfrak{D}}_g,\ {\mathcal{G}}_{en\ \left(g\right)}$ in each UAV \textit{g} have their impact on the computation discriminator part will enhance the result and on the other hand generator will minimize the result of the value function.

Previous work proposed in \cite{zhang2021distributed},\cite{liu2021lstm} supports nash equilibrium (NE) maintained by autonomous CGAN model given that learned distribution function and discriminators satisfy the condition $b^{{\mathcal{G}}_{en(g)}}_g=b_g$and ${\mathfrak{D}}_g=0.5$ such that the learned distribution of generator is same as dataset distribution, given that discriminator value is at half so that model cannot differentiate between real data and generators data. But in the actual case where resources are limited therefore the model is bound to train on restricted length of dataset sample array acquired from the channel. Self-sustained CGAN can learn the spectrum of the assigned path concerning AoA and AoD and the model works accurately within the same operating region. However, in a diverse environment, where the operating region is dynamically changing i.e. UAV changes its path or GS mobility are some factors that require each model training recursively through transmitting pilot symbols in to and fro manner. Due to this, initially, the channel modeling process is not much challenging but updating a channel with time is a resource and time consuming for the airborne network. To overcome this challenge (that no single dataset collected by each UAV overlaps with the dataset of others) one resource-efficient scheme can be introducing a real-time cooperative information sharing method in our CGAN model. But exchanging real-time data within UAVs can also bring a lot of communication overhead by consuming spectrum resource heavily \cite{zhang2021distributed}.

Also, another idea is instructing a UAV to share its location with a timestamp with others, but this scheme may bring privacy issues because within mmWave spectrum there may exist multiple network providers, and each have their own GS and UAVs. In consideration of all these pros and cons, there is a need to have an efficient, less resource consuming, and privacy-preserving data sharing method of A2G link in the mmWave channel model.

\subsection{LSTM-DCGANs Framework}

Nowadays, distributed network for the generative adversarial model is one of the hot areas of study in machine learning. Previously, machine learning shows a major contribution in the vision-based application, subsequently in baseband signal processing. However, specifically implementing LSTM-DCGAN in real-life problems has a very limited application \cite{ferdowsi2019generative} but till now no one has applied the LSTM-DCGAN model in wireless communications especially in A2G mmWave channel estimation. \cite{zhang2021distributed} did used Simple DCGAN for modeling channel for UAVs but that work lacks in providing required latency and convergence. Moreover, it was not suitable of maritime UAVs networks. 

Therefore, to fill this research gap, we here explain one of our distributed CGAN idea, in which generative channel models learn from each dataset collected from a group of samples acquired by UAVs operating in a distributed manner. Keeping in mind the data privacy issue, we proposed a method which works by collaborating without sharing actual raw data. This is similar approach as followed by \cite{zhang2021distributed} but we have improved this approach by considering spatio-time features. We say that set of UAVs \textit{G }have collected \textit{g} datasets and final model is developed by taking union of all individual dataset. Thus, we denoted by $H=\ H_1\cup H_2\cup ,,,,,,,H_g$, where \textit{H }is local dataset of channel sampled from several location and each distribution function $b_g$ only meant for specific geographic area, for overall spatial space modeling we need to consider all datasets of all areas.

Ideally an efficient generator distribution $b^{{\mathcal{G}}_{en}}_g$is trained by each airborne object \textit{g} to obtain generic distribution \textit{b} of network channel in a condition such that no UAV \textit{g} will share any actual dataset $H_g.$ For this, in our scenario of channel estimation for 5G maritime UAV, we merged the idea of distributed brainstorming GANs proposed in \cite{ye2020deep} with channel modeling through LSTM-DCGAN. Furthermore, in our concept of distribution, each airborne object \textit{g }cooperates with others by sharing AoA and AoD conditions along with generated samples ${\mathcal{G}}_{en\ (g)}$ instead of raw information.  

Fig.2 is a basic illustration of our proposed idea, where we assumed \textit{g }UAVs within the network each airborne object comprises of a generator, condition, discriminator, and collected dataset. Each discriminator has three own inputs and one shared input, among three own inputs one is from the generator, the second is from the dataset and the third input is a defined condition. Whereas, shared input is provided by the generator of the adjacent UAV. Here it is worth noting that in our proposal instead of sharing real datasets, we are sharing generated data. Therefore, for fooling discriminators all generators cooperatively generate data for sharing and each will try to differentiate between actual data and generated data.  To elaborate this concept, we denote ${\mathcal{C}}_g$ as a UAV sending generated dataset to$\ g$\textit{ }UAV, whereas $Q_g$be a variable to represent a UAV to whom $gth\ $UAV is sending its generated dataset. Communication between the generator and discriminator of each UAV can be represented by the game-theoretic model as in \eqref{7}.
\begin{equation}\label{7} 
\begin{split}
R_{g}(\mathfrak{D}_{\mathfrak{g}},{\mathcal{G}}_{en\ (g)},{\mathcal{G}}_{en\ (n)_{{n\epsilon C_g}}})=\frac{1}{\mathcal{I}}\sum_{\mathcal{I}=1}^{\mathcal{I}}\varepsilon_{h\sim b_{g}^{b}}[\log{\mathfrak{D}_{g}}(h\mid\varphi_{\mathcal{I}})]+\\ \varepsilon_{{\mathfrak{z}} \sim b_{g}^{\mathfrak{z}}}[\log(1-\mathfrak{D}_{\mathfrak{g}}({\mathcal{G}}_{en\ (g)}({\mathfrak{z}}\mid\varphi_{\mathcal{I}})]
\end{split}
\end{equation}

The joint distribution of dataset from \textit{g }UAV and dataset sample from all adjacent UAVs set $\mathcal{C}$ are denoted by: $b^b_g={\pi }_gb_g+\sum_{j\epsilon C_g}{{\pi }_{gj}b^{{\mathcal{G}}_{en}}_j}.$  In joint distribution, we referred ${\pi }_i=\frac{H_g}{H_g+\eta \sum_{j\epsilon C_g}{H_j}}$ and ${\pi }_{ij}=\frac{{\eta H}_g}{H_g+\eta \sum_{j\epsilon C_g}{H_j}}$ respectively, whereas,  channel samples shared by UAV $j$\textit{ }to UAV $g$\textit{ } in each time frame are denoted by-product of $\eta $ with a dataset of UAV (${\eta H}_g$) with condition that $\eta $ will always greater than 0. Unlike in \cite{ye2020deep} and \cite{zhang2021distributed} the value functions are linearly dependent on each other, for this we model distributed UAV network mathematically as,

\begin{align} \label{8} 
& R(\ \{{\mathfrak{D}}_g{\}}^{{\mathcal{G}}_{en\left(g\right)}}_{g=1},\{{\mathcal{G}}_{en\left(g\right)}\}\ {\ }^{{\mathcal{G}}_{en\left(g\right)}}_{g=1})\\ &=\sum^G_{g=1}{R_g({\mathfrak{D}}_g,{\mathcal{G}}_{en\left(g\right)},\ \left\{{\mathcal{G}}_{en\left(j\right)}\right\}{\ }_{j\epsilon C_g})} \nonumber
\end{align} 

From \eqref{8} it can be inferred that all discriminators coerce to maximize the value function, on the other hand, all generators will try to minimize the utility function. Hence, appropriate discriminators and generators for distributed CGAN training can be represented as minima and maxima in \eqref{9}.

\begin{equation} \label{9} 
{\{{\mathfrak{D}}^*_g\}}^G_{g=1},\ {{{\{\mathcal{G}}_{en}}^*_g\}}^G_{g=1}={arg}^{min}_{{\mathcal{G}}_{en(1)},\dots {\mathcal{G}}_{en(G)}}\ \ {arg}^{max}_{{\mathfrak{D}}_1,\dots {\mathfrak{D}}_G}\ R 
\end{equation} 

The architecture of 5G enabled UAVs operating in the network will decide LSTM based discriminator and generator values. Previously work in \cite{ye2020deep} and \cite{goodfellow2014generative} proposed exploiting distributed GAN but without data sharing or network optimization for data sharing. Similarly \cite{zhang2021distributed} only considered simple DCGAN approach. In this work we present an optimal network structure for 5G enabled maritime UAVs with  optimized learning solution.

\subsection{LSTM-DCGAN Coupling}

We use a graph $\mathbb{G}=\left(G,\ \mathbb{E}\right)$ to represent UAVs on the communication network, where a set of UAVs is represented by \textit{G }and $\mathbb{E}$ to denote a set of edges. We further say that ${\mathrm{e}}_{gj}\epsilon \mathbb{E},$ each edge is obtained from UAV pair communicating over the air to air mmWave link. For clarity let us say, within each loop iteration of CGAN learning, airborne object \textit{g} will share its generative data with the discriminator of \textit{j }UAV. We call in-degree here to a UAV receiving generated data from $gth$\textit{ }UAV at ${\mathcal{C}}_g=|{\mathcal{C}}_g|$ and out-degree is an airborne object to whom UAV \textit{g }is sending its data at $Q_g=|Q_g${\textbar}. 

In parallel, we define a route between points \textit{x }and \textit{y }covered by UAVs such that traversing on edge $\mathbb{E}$ from point \textit{x} to \textit{y} in joined and non-overlapping manner. Whereas, length of the covered path ${\ell }_{x,y}$should be equivalent to the total number of edges over the path ${\mathbb{E}}_{x,y}$, and for covered loop, having the same starting and ending point i.e. \textit{x} we use the notation ${\mathbb{E}}_x$with total loop length expressed as ${\ell }_x$.

For air to air (A2A) communication study proposed in \cite{bas2017real} has utilized orthogonal frequency divisional multiplexing (OFDM) resource chunks (RC, mathematically $\mathfrak{B}$ and $\mathfrak{B}\ge G\ nummber\ of\ UAVs$) for channel dataset produced by generator at link sub 6GHz. Therefore, for lossless communication and to reject cross talk within UAV topology, we prefer to have number resource blocks always greater than or equal to communication channels (i.e. $|\mathbb{E}|\le \mathfrak{B}$). 

Thus, we define transmission between UAV \textit{g }to UAV \textit{j }consuming RC block $b$ have communication rate ${C\_R}_{gj}={\omega }_b{log}_2(1+\frac{P_{w\left(g,j\right)}{{P\_}_l}_{\left(gj\right)}}{{\sigma }^2})$, here omega (${\omega }_b)$  and ${\sigma }^2$ represents covered bandwidth and noise power, whereas, $P_{w\left(g,j\right)}\ $and ${{P\_l}_{(gj)}}_{\ }$denote signal power and path loss respectively. For lossless A2A communication with optimal SNR ($S_{NR}$) between UAV pair \textit{g }and \textit{j, }no resource block RC will be consumed if at UAV \textit{j} received SNR is lower than the threshold $T_{Th}$ ($\frac{P_{w\left(g,j\right)}{{P\_}_l}_{\left(gj\right)}}{{\sigma }^2}\ <\ T_{Th}$). Finally, for distribution, each airborne object \textit{g} is\textit{ }restricted to shares its generated dataset with adjacent flying object $Q_g$ within stipulated transmission time slot $t_s$.

Here, for LSTM-DCGAN model learning time can be estimated by counting the total number of iterations required in complete learning along with benchmarking of each iteration, we say this as convergence time of the model and denote it as $t_{con}$. For analysis, we assumed our network is homogenous and have fixed datasets from each airborne object $H=H_1,\dots,H_G$ and for synchronized learning, UAVs transmitting dataset to \textit{g }UAVs, we assumed $\mathcal{C}={\mathcal{C}}_1,\dots,{\mathcal{C}}_G$. Then, probability of convergence for learning within $I_{tr}$ number of iterations is given in lemma 1.

\subsubsection{Lemma 1}

\textit{Probability }$P_{rob(G)}(I_{tr})$\textit{ of covering whole distribution b by each g UAV's LSTM-generator distribution }$b^{{\mathcal{G}}_{en}}_g$\textit{ within }$I_{tr}$\textit{ iterations by UAV network topology }$\mathbb{G}$\textit{ with a defined upper limit in training error }$T_{error}$\textit{ is given by }\eqref{10}:

\begin{strip}
\begin{equation}\label{10}
P_{rob\left(G\right)}\left(I_{tr}\right)=\begin{cases}0 & 0<I_{tr}<\ell_{max}\\\frac{\left[\left(1-T_{Error}\right)\eta\right]^{l_{max}}}{C}{\left(1+C\eta\right)^{l_{max}}} & I_{tr}=l_{max} \\ P_{rob(G)}(l_{max})+\sum_{g=l_{max}+1}^{I_{tr}}[\prod_{j=l_{max}}^{g-1}1-\frac{\left[\left(1-T_{Error}\right)\eta\right]^{l_{max}}}{\left(1+C\eta\right)^{j-1}}] \frac{\left[\left(1-T_{Error}\right)\eta\right]^{l_{max}}}{\left(1+C\eta\right)^{g-1}}& l_{max}<I_{tr}<\ l_{max}+l_{loop}^{min}
\end{cases}
\end{equation}
\end{strip}

$and\ for\ I_{tr}>{\ell }_{max}+{\ell }^{min}_{loop}$

Here we represented the length ${\ell }_{max\ }$ as the maximum shortest path between \textit{x }and \textit{y} in graph $G$ and denote ${\ell }^{min}_{loop}$ for shortest loop path with the same starting point. Whereas, $\gamma \left(I_{tr}\right)\ge 1$ defined for the coefficient of acceleration.

\begin{strip}
\begin{equation}\label{11}
\begin{split}
P_{rob\left(G\right)}\left(I_{tr}\right)=P_{rob\left(G\right)}\left({\ell }_{max}+{\ell }^{min}_{loop}-1\right)+\sum^{I_{tr}}_{g={\ell }_{max}+{\ell }^{min}_{loop}}{[\prod^{g-1}_{j={{\ell }_{max}+\ell }^{min}_{loop}-1}{(}}1-\frac{[(1-T_{Error})\eta ]^{{\ell }_{max}}}{(1+\mathcal{C}\eta )^{j-1}}\prod^j_{\mathcal{I}={{\ell }_{max}+\ell }^{min}_{loop}-1}\\{\gamma (\mathcal{I}))]}\frac{[(1-T_{Error})\eta ]^{{\ell }_{max}}}{(1+\mathcal{C}\eta )^{g-1}}\prod^j_{\ell ={{\ell }_{max}+\ell }^{min}_{loop}}{\gamma \left(\ell \right)}
\end{split}
\end{equation}
\end{strip}

Lemma 1 states three distinct operating conditions, two for convergence and one for do nothing. First to maximize the convergence process number of iterations should always be greater or equal to the maxima of the shortest path length in network graph $\mathbb{G}$. Otherwise, if the number of iterations is less than the maximum of the shortest path length then there will be no convergence. Hence, to increase the probability of convergence for dataset sharing within the UAV network, it is mandatory to mitigate maximum length such that it becomes less than the number of required iterations. Probability for LSTM-discriminator ($P_{rob(\mathfrak{D})}(I_{tr})$) is equal to probability of LSTM-generator distribution. Thus we can say that $P_{rob(G)}(I_{tr})$=$P_{rob(\mathfrak{D})}(I_{tr})$. Subsequently in the next stage, if we relate the same concept in LSTM-DCGAN model, then the model will converge with probability $P_{{rob}_{LSTM-DCGAN}}$ on $I_{{tr}_{LSTM-DCGAN}}\epsilon {\mathcal{N}}^+$ the number of iterations expressed in \eqref{12}.

\begin{equation} \label{12} 
P_{rob\left(G\right)}\left(I_{tr\left(G\right)}-1\right)< P_{{rob}_{LSTM-DCGAN}}\le P_{rob\left(G\right)}\left(I_{tr\left(G\right)}\right) 
\end{equation} 

It is now clear from \eqref{10} that with probability the iterations required by the generator of each airborne object to fully adopt with complete channel distribution will be minimum as compared to the \cite{zhang2021distributed}. For analysis, we assumed to include fix dataset, therefore, for training LSTM-DCGAN's generator and discriminator locally, $T_{error}$ is the upper limit of training error that an object attained in a fixed timestamp $t_{fx}$. 

Now, for network graph $G$, the convergence time required for learning distributed LSTM-DCGAN is expressed in \eqref{13}.

\begin{equation} \label{13} 
C_{NVRG}(G,\mathfrak{D})= (t_{T_{Th}}+ t_{fx}). I_{tr\left(G\right)}I_{tr\left(\mathfrak{D}\right)})
\end{equation} 

Therefore, for designing an optimal channel model of A2A link among distributed collaborative airborne objects in network $G$ with the constraint of limited resources, the goal is to optimize the network such that convergence time of LSTM-DCGAN learning process over each UAV is minimized. Mathematically this optimization can be expressed as:
\begin{align}\label{14}
& \min_{G,\mathfrak{D}}C_{NVRG}(G,\mathfrak{D}) \\ s.t. \quad
&\sum_{t_{error(g,j)\epsilon T_{error}}}P_{w(g,j)}\leq P_{w(max)} , \forall gj\epsilon G,\mathfrak{D} \nonumber \\ & \frac{P_{w\left(g,j\right)}{P_l}_{gj}}{P_{Sym}}{\sigma^2}\geq I_{tr},\forall c_{gj}\epsilon C_{link} \nonumber \\
&\frac{\eta H_g\rho}{{C_R}_{gj}}\le t_{T_{Th}}\ \forall c_{gj}\epsilon C_{link} \nonumber\\& \exists T_{error(gj)}\subset C_{link} \forall gj\epsilon G \nonumber \\ & G\le\left|C_{link}\right|\le\mathfrak{D} \nonumber
\end{align}

\eqref{14} describes obtaining global minima of convergence time from the graph of network $\mathbb{G}$ for LSTM generator $G$ and LSTM discriminator $\mathfrak{D}$ in a condition defined when in A2A link between UAVs from \textit{g} and \textit{j} are communicating with each other with transmit power $P_{w(g,j)}$ which should be less than the maximum transmission power $P_{w(max)}$. Next the threshold is defined for optimal SNR in A2A link, which is obtained by having a product of transmitting power between the object from two groups with respective path loss affected by noise power. In other conditions defined in \eqref{14}, LSTM-DCGAN convergence time is formulated by-product of the number of samples and size of acquired data sample by operating UAV \textit{g.}  

Lastly the learning from tightly couple dataset sharing within CGAN network and interference rejection within link respectively. In the given scenario, our objective is to find the minimum convergence time of path between any two airborne objects operating within the optimized graph ${G}^'\ $from available network $G$ instead of sorting shortest path inside network $G$. 

Apparently, for solving \eqref{14} in all, overall a  the controller need to defined conditions for network optimization depending upon path loss between pair of 5G enabled UAVs. But involving a controller in the distributed network will change the decentralized structure into centralized, which is not acceptable. Therefore, to address this challenge, in the following section we devise a method to decimate \eqref{14} into small conditions handled by each UAV. Afterward, with distributed learning in \eqref{9}, we will formulate Nash Equilibrium.  

\subsubsection{Proof of Lemma 1}
In order to support our derivation and results for the convergence rate of distributed learning model, we assumed samples shared in each iteration by each generator are independent of others in the network. Here we have followed similar approach as of \cite{zhang2021distributed}. Here, each generated sample from the respective generator has a similar amount of CSI. Therefore, in the context of CSI from the locally available dataset, each collected dataset can be defined as an independent and identically distributed random process and we further used recursion to prove Lemma 1.  Suppose $\{g=x_1,\dots \dots ..x_{l_{max}}\}$ is the set of UAVs defined over global maxima of shortest path and we considered the portion of CSI whose size is equivalent to the information contained inside a single dataset sample.

\textbf{A Single Time Probability }
Initially, we calculated the probability of sharing the same real-time collected CSI (whose size is equal to the size of a single dataset) from $H_g$ to the generator of an airborne object $x_{{\ell }^{max}}\ $at iteration $I_{tr}$.

\begin{enumerate}
\item  For $I_{re}<{\ell }^{max}:$ Within each loop iteration any channel sample can be shared with the adjacent UAV. Therefore, when iterations are less than the maxima of the shortest path no information sharing can occur between \textit{g }to $x_{{\ell }^{max}}$ UAVs via the shortest path. Hence the probability of information sharing in this condition is zero.

\item  For ${\ell }^{max}\le I_{tr}<{\ell }^{max}+{\ell }^{min}_{loop}$: Within the first iteration of the simulation $I_{tr}=1$, airborne object \textit{g} will transmit generated channel dataset $\eta H$ to the UAV $x_1$. Therefore, the probability of event transmitting portion of considered CSI from UAV \textit{g }to $x_1$ is equivalent to the sampling ratio given by $\eta $. On the other hand, if the probability of training error occurring at the local generator is $T_{error}$, then the probability of error-free training is given by $1-T_{error}$. Hence, the probability of transmitting portion of CSI from \textit{g }UAV to $x_1$is given by $P_{rob}{\ }^{in}_1=(1-T_{error})\eta $. In the meantime, UAV $x_1$ receive samples shared by the generator of other operating objects from neighbors $C-1$ from set $C_{x1}$, therefore, the contribution ratio of UAV \textit{g'}s information in the overall dataset of $x_1$UAV is calculated by $P_{rob}{\ }^{out}_1=\frac{P_{rob}{\ }^{in}_1}{1+C\eta }=\frac{(1-T_{error})\eta }{1+C\eta }$. Now in the second iteration $I_{tr}$=2, samples $\eta H$ shared by the generator of $x_1$ UAV with the UAV $x_2$. Then the probability of moving \textit{g}'s information from $x_1$ UAV to $x_2$ UAV is calculated by:

$P_{rob}{\ }^{in}_2=\eta P_{rob}{\ }^{out}_1$=$\frac{1-T_{error})\eta ]{\ }^2}{1+C\eta }$. Also, the percentage of $gth$ UAV's information is reduced in size on reaching at UAV $x_2$ because of data generation from other airborne objects, hence, we denote the probability of data at $x_2$UAV as $P_{rob}{\ }^{out}_2=\frac{P_{rob}{\ }^{in}_2}{1+C\eta }=\frac{[\left(1-T_{error}\right)\eta ]{\ }^2}{(1+C\eta ){\ }^2}$. This procedure continues recursively till complete delivery of sample information at UAV $x_{{\ell }^{max}}$ at iteration $I_{tr}={\ell }^{max}$. 

Therefore, the probability of reaching $gth$ UAV's data at $x_{{\ell }^{max}}$ UAV in ${\ell }^{max}$ iterations are calculated by $P_{rob}{\ }^{in}_{{\ell }_{max}}=\frac{[\left(1-T_{error}\right)\eta ]{\ }^{{\ell }^{max}}}{(1+C\eta ){\ }^{{\ell }^{max}-1}}$. From the above discussion, we can conclude that, like the FIFO concept, former data samples held at each UAV get reduced in size by $\frac{1}{1+C\eta }$ because of the latest incoming generated data from \textit{C }adjacent operational UAVs. In the meantime, within each jump of data sharing among UAVs on the path, data size is reduced depending upon training error (1-$T_{error})\eta \ $and sampling ratio. Hence, we can say that probability of successfully receiving the shared portion of CSI at $x_{{\ell }^{max}}$ UAV in $I_{tr}$ is given by $P_{rob}{\ }^{in}_{{\ell }^{max}}(I_{tr})=\frac{[\left(1-T_{error}\right)\eta ]{\ }^{{\ell }^{max}}}{(1+C\eta ){\ }^{{\ell }^{max}-1}}$.

\item  For $I_{tr}\ge {\ell }^{max}+{\ell }^{:min}_{loop}$, In the condition when iteration is greater than the length of shortest path of the ring with $gth$ UAV, information of its own distributed data will start arriving from the adjacent object $C_g$\textit{ }connected in the ring loop. Therefore, data size reduction, in this case, is comparatively higher than $\frac{1}{1+\mathcal{C}\eta }$. Hence, the acceleration coefficient should be greater than 1 ( $\gamma \left(I_{tr}\right)>1)$, and 1+$C\eta $ will be added to reduction factor when ( $\gamma \left(I_{tr}\longrightarrow +\infty \right)$. Thus, we denote the probability of data distribution as:

${P_{rob}}^{in}_{{l}_{max}}\left(I_{tr}\right)=\frac{[\left(1-I_{tr}\right)\eta ]{\ }^{{\ell }^{max}}}{(1+C\eta ){\ }^{I_{tr}-1}}\prod^{I_{tr}}_{g={\ell }^{max}+{\ell }^{min}_{loop}}\\{\gamma \left(g\right)for\ I_{tr}\ge {\ell }^{min}_{loop}}$.
\end{enumerate}

\textbf{Cumulative Probability }
Here we will describe the cumulative probability that portion of $gth$ CSI is successfully arrived at $x_{{\ell }^{max}}$ UAV after $I_{tr}$ number of iterations

\begin{enumerate}
\item  $I_{tr}<{\ell }^{max}:$ Provided that ${P_{prob}}^{in}_{{\ell }^{max}}\left(I_{tr}\right)=0$, then the probability of general distribution is $P_{rob\left(G\right)}\left(I_{tr}\right)=0\ for\ all\ I_{tr}<{\ell }^{max}$.

\item  $I_{tr}={\ell }^{max}:$ when iterations are exactly equal to length then probability will be $P_{rob\left(G\right)}\left(I_{tr}\right)=\\{P_{rob}}^{in}_{{\ell }^{max}}\left({\ell }^{max}\right)=\frac{[\left(1-I_{tr}\right)\eta ]{\ }^{{\ell }^{max}}}{(1+C\eta ){\ }^{{{\ell }^{max}}_{\ }-1}}$.

\item  ${\ell }^{max}<I_{tr}<{\ell }^{max}+{\ell }^{min}_{loop}:$ in this condition probability of complete information sharing is formulated by chain rule, such that:

$P_{rob\left(G\right)}\left(I_{tr}\right)={P_{rob}}^{in}_{{\ell }^{max}}$(${\ell }^{max})$+[1-${P_{rob}}^{in}_{{\ell }^{max}}$(${\ell }^{max})$]\\${P_{rob}}^{in}_{{\ell }^{max}}\left({\ell }^{max}+1\right)+ \dots \dots +\prod^{I_{tr}-1}_{g={\ell }^{max}}\\{\left[1-{P_{rob}}^{in}_{{\ell }^{max}}\left(g\right)\right]{P_{rob}}^{in}_{{\ell }^{max}}\left(I_{tr}\right)}$. 

Finally, we can write:

$P_{rob\left(G\right)}\left(I_{tr}\right)={P_{rob}}^{\ }_{{(G)}^{\ }}$(${\ell }^{max})$ +$\sum^{I_{tr}}_{g={\ell }^{max}+1}\\{[\prod^{g-1}_{j={\ell }^{max}}{(1-\frac{[(1-T_{error})\eta {]\ell }^{max}}{(1+C\eta )^{j-1}})]\frac{[(1-T_{error})\eta {]\ell }^{max}}{(1+C\eta )^{g-1}})]}}$

\item  $I_{tr}\ge \ {\ell }^{max}+{\ell }^{min}_{loop}:$ Inn condition when data sharing start inside the loop, the probability is as follows:

$P_{rob}\left(I_{tr}\right)=P_{rob\left(G\right)}\left({\ell }^{max}+{\ell }^{min}_{loop}-1\right)+\sum^{I_{tr}}_{g={\ell }^{max}+{\ell }^{min}_{loop}}\\{[\prod^{g-1}_{j={\ell }^{max}+{\ell }^{min}_{loop}-1}{(1-\frac{[(1-T_{error})\eta ]^{{\ell }^{max}}}{(1+C\eta )^{j-1}}\prod^j_{k=\ {\ell }^{max}+\\{\ell }^{min}_{loop}-1}{\gamma (k))]}}}\ \\ \frac{[(1-I_{tr})\eta ]^{{\ell }^{max}}}{(1+C\eta )^{g-1}}\prod^g_{\ell =\ {\ell }^{max}+{\ell }^{min}_{loop}}{\gamma (\ell )}$ 

we further assume that $\gamma \left(\ {\ell }^{max}+{\ell }^{min}_{loop}-1\right)=1$. 
\end{enumerate}

\subsection{LSTM-DCGAN Learning Optimization}
For LSTM-DCGAN network, in this section, we will optimize \eqref{8}, \eqref{9} and \eqref{14} jointly for more precise channel estimation in airborne communication, by initially deriving the most feasible network topology denoted by ${G}^'$. Secondly, using derived ${G}^'$.structure, for each airborne 5G enabled UAV \textit{g}, we will logically formulate the most suitable LSTM-DCGAN model (${\mathcal{G}}^'_{en},\ {\mathfrak{D}}^'_{\ })$.

For optimizing \eqref{14} in a decentralized manner by eliminating the server, we converted two inequalities of into joint conditions with equality and inequality as in \eqref{15}. Such that initially to achieve distribution, the number of A2A communication links should be equivalent to the number of operating UAVs that are consuming fewer resources than the number of links. Furthermore, considering \eqref{14}, we derive the ideal 5G enabled UAV network in Lemma 2.
\begin{equation} \label{15} 
G=\left|C_{link}\right|\le \mathfrak{B}=\min({\mathcal{G}}^'_{en},\ {\mathfrak{D}}^'), 
\end{equation} 

\subsection{Lemma 2}
\textit{The restriction defined in \eqref{15} illustrates that 5G enabled UAVs should be tightly coupled with each other inside the network forming ring topology.} Such that, $C_g=Q_g=1,\ C_g\cap Q_j={\emptyset }_{\ ,\ }and\ Q_g\cap Q_j=\ \emptyset ,\ {\forall }_{g,j}\epsilon G\ and\ g\neq j$.

\textbf{Lemma 2 Proof:}

We know that our topology is a tightly coupled ring network graph, where each corner has a minimum of one incoming angle and an outgoing angle.  Therefore, following expression in \eqref{14} tightly coupled connection requires each node to have in-out edges, at  $C_g\ge 1$ and $Q_g\ge 1$. Also, it is given in equation \eqref{15} that total entailed edges are equivalent to the 5G connected UAVs in-network, mathematically, $\sum^G_{g=1}{C_g=G\ \mathrm{and\ }\sum^G_{g=1}{Q_g=G\ }}$. Hence, we can prove that $C_g=$ $Q_g=1,\ {\forall }_g\in G.$ Therefore, with only one in-out edge available at each UAV, it is proved that the network is coupled in ring topology such that $C_g\bigcap C_j$ is Empty set and $O_g\bigcap O_j$ is also empty\ set where $g\neq j$.  

The optimization constraint mentioned in \eqref{14} and \eqref{15} requires a communication system of UAV operating in-ring network, such that each UAV is receiving channel sample dataset from one who is sharing it and subsequently this UAV is sharing its own generated dataset with the adjacent one to form a ring. Therefore, optimization objective based on the above two theorems, we convert our centralized \eqref{14} in to distributed and decentralized optimized expression handled by each UAV individually to reduce convergence time with the maximum value of shortest path only when UAV from set \textit{g }shares its channel dataset with the UAV of group \textit{Q }( which lies in position giving minimum convergence time). Hence, \eqref{14} can be re-written in distribution as follows:
\begin{align} \label{16} 
& \min_{q_{i}\epsilon G_{-g}}{{\ell }^{max}_{g}}(G.\mathbb{E}_{g,q_g})  \\
 s.t. \quad & P_{w_{g,q_{g}}}\le \ P_{w(max)}, \nonumber\\
& P_{w_{g,q_{g}}}{P\_{l}}_{{g,q}_{g}}{\sigma }^2\ge \ T_{Th}, \nonumber\\ 
&\eta H_{g}\rho /{C\_{R}}_{g,q_g}\le t_{T_{Th}} \nonumber
\end{align}

In this the $G_{-g}$ represents a group of airborne objects in set \textit{G} excluding $gth$\textit{ }UAV. Whereas, after adding an edge ${\mathbb{E}}_{g,q_g}$in our network of UAVs $G$, we denoted${\ \ \ \ell }^{max}_g$ as global maxima of the shortest path between airborne object \textit{g }to the other. Furthermore, we defined a set of all adjacent UAVs $\mathfrak{S}$ with whom \textit{g }can collaborate in sharing dataset under conditions expressed in \eqref{16}. 

\begin{equation} \label{17}
\begin{split} 
{\mathfrak{S}}_g=\{j\epsilon G_{-g}|P_{w(gj)}\le P_{w({\mathrm{max} \ \ }\mathrm{)}}. P_{w_{g,q_g}}{P\_l}_{{g,q}_g}{\sigma }^2 \ge \ T_{Th}, \\ \eta H_g\rho /{C_R}_{g,q_g}\le t_{T_{Th}}\} 
\end{split}
\end{equation}

Based on the distributed equation in \eqref{16} the mandatory conditions are:

\textbf{Proposition 1 }(Necessary Condition)
\textit{For constraint in \eqref{15}, feasible 5G enabled UAV network topology }${G}^'$\textit{only exist on holding condition }${\bigcup }^G_{g=1}{\mathfrak{S}}_g=G\ where,\ {\forall }_g,\ {\mathfrak{S}}_g\neq \emptyset $\textit{}

\textbf{Proof of Proposition 1}

\noindent Boundary conditions required in proposition 1 are proved using counter-arguments. First of all, we suppose that each UAV denoted by \textit{g }belongs to set \textit{G,} when ${\mathfrak{S}}_g$ set of feasible UAVs is zero. Further, we say $Q_g=0$, which is against the requirement of a tightly coupled network with $Q_g\ge 1.$ Then, if we say for each feasible adjacent UAV which is a subset of \textit{G }( $\sum^G_{g=1}{{\mathfrak{S}}_g\subset G})$\textit{, }then there is at least one UAV in set $g\ $, such that no other airborne is bound to share any generated dataset with it. Hence, for $gth$\textit{ }UAV, $C_g$is zero, which is again counter to the requirement of a strongly connected network with $C_g\ge 1.$ Thus, we conclude that strong coupled connection for optimal UAV topology ${\mathbb{G}}^*$ union condition (${\bigcup }^G_{g=1}{\mathfrak{S}}_g=I\ and\ {\forall }_g,{\mathfrak{S}}_g\neq \emptyset )$ must be true.

Strongly coupled connections are not possible among 5G enabled UAVs operating within a network if the union of the optimal set of 5G connected UAVs does not cover all airborne objects. Hence, in this condition, no optimal solution exists for \eqref{16}. Following Lemma 2 and Proposition 1, we formulated some below-mentioned conditions to attain optimal network topology.

\textbf{Proposition 2 }(Sufficient Condition)
\noindent Constraint given in \eqref{15} and if union set holds \textit{ }${\bigcup }^G_{g=1}{\mathfrak{S}}_g=G\ where,\ {\forall }_g,\ {\mathfrak{S}}_g\neq \emptyset $ for every value of \textit{g}, the most optimal airborne network is given by  ${G}^'=(G,\mathbb{E})$. Whereas, we know that $\mathbb{E}\subseteq \{{\mathbb{E}}_{g,j}|g\epsilon G,j\epsilon {\mathfrak{S}}_g\}$ and ${G}^'=G-1,$ ${\forall }_g\epsilon G.$

\textbf{Proof of Proposition 2}

\noindent Here we will describe how proposition 2 is more feasible for equations \eqref{14} and \eqref{16} then we will define its suitability for expression \eqref{16}. Initially, we know that all edges inside a network are formed by UAVs of set \textit{G }and set of feasible $\mathfrak{S}$ ($\mathbb{E}\ \subseteq \left\{{\mathbb{E}}_{gj}\right|g\ \epsilon G,\ j\epsilon {\mathfrak{S}}_i\}$ ). Hence, it is clear that constrictions defined in \eqref{14} and \eqref{16} are satisfied as per the basis of feasible set in \eqref{17}. Furthermore, it is known that the maxima of the shortest path in a feasible UAV network should be one less than the total number of UAVs in \textit{G, }mathematically, ${\ell }^{max}_g\left({G}^*\right)=G-1$, this also leads to having ring-based network topology in which UAVs are strongly coupled with each other as per constraint in equation \eqref{14}. Also, this constraint leads to maintain an equal number of communication links with UAVs holding equation \eqref{14}. Hence, we proved our feasible solution here. From lemma 2, we can only form a ring topology for airborne UAV network only when ${\ell }^{max}({G}^*)$ has a constant value of \textit{G-1. }Hence, with the fixed value of ${\ell }^{max}({G}^*)$, convergence will also be fixed with any given sequence of ring topology which yields a similar convergence time for all elements of the set.

Constraints mentioned in \eqref{14} and \eqref{15}, a communication network link among edges $\mathbb{E}\subseteq \left\{{\mathbb{E}}_{gj}\right\}g\epsilon G,j\epsilon {\mathfrak{S}}_g$, where convergence time of connected nodes in a ring topology is minimum is defined as an optimal network ${G}^'$. In proof of proposition it is elaborated that optimal network ${G}^'$ is not the only solution to the challenge in \eqref{16}, however, it also provides the most suitable idea for decentralized computing over distributed computing of equation \eqref{14}. Here one derivation that is important to note is, proposition 2 require having $Q_g=1$, on the contrary to the equation \eqref{14}, states that $Q_g$ should be greater than one or edges count should be greater than or equal to the number of airborne objects, provided that number of edges formed by 5G enabled UAVs communicating within network $|\epsilon |$should hold equality with $\sum_{g\epsilon G}{Q_g\le }\mathfrak{B}$. Now we will formulate an optimized network architecture with condition $Q_g\ge 1.$

\noindent We already devised a feasible network ${G}^'$, now to achieve tightly coupled connection property from equation \eqref{14}, we are taking reference optimal network structure ${G}^'$, where we will more edges to already feasible network ${G}^'$ to reduce convergence time $C_{NVRG}(G)$. Therefore, likewise in \eqref{17}, we illustrate a similar feasible group set for UAVs \textit{g }which fulfill constraints in \eqref{14} such that:
\begin{align} \label{18} 
& \widehat{\mathfrak{S}_{g}}={j\epsilon} {\mathfrak{S}}_{g} \ at\ \sum_{j\epsilon {\mathfrak{S}}^{q}_{g}}P_{w(gj)}\le P_{w(max)},\\
&\frac{P_{w(gj)}P_{lgj}}2{{\sigma }^{2}}{\ge T_{Th},\ \eta H_{g}\rho /{C\_{R}}_{gj}\le t_{T_{Th}}} \nonumber
\end{align} 

We assume that $\widehat{{\mathfrak{S}}^q_g}\subseteq \widehat{{\mathfrak{S}}_g}\subseteq {\mathfrak{S}}_g$, such that subset ${\mathfrak{S}}^q_g$contains elements from the set $Q_g$which are also elements of ${\mathfrak{S}}_g$.For feasible network where $Q_g=1,\ {\mathfrak{S}}_g=\widehat{{\mathfrak{S}}_g}=\widehat{{\mathfrak{S}}^q_g}$. Therefore, for feasible topology defined in \eqref{18}, the following corollary will express the optimal solution for \eqref{14}.

\noindent \textbf{Corollary 1 (}Necessary Condition\textit{), optimal network architecture denoted by }${G}^*=(G,\ \mathbb{E}$\textit{), which maximize convergence rate in \eqref{14}, such that feasible network becomes a subset of optimal network and }$\mathbb{E}=\left\{{\mathbb{E}}_{gj}\right|{\forall }_g\epsilon G,\ {\forall }_j\epsilon $\textit{ }$\widehat{{\mathfrak{S}}^q_g}$\textit{$\}$.}

\textbf{Proof of Corollary 1}

\noindent In proposition 1, we proved ${G}^'$as the feasible network graph fulfilling the limitation of the tightly coupled network. Therefore, belonging of optimal network to feasible network condition (${G}^'\subseteq \ {G}^*)$ will certify that ${G}^*$ also, follow the constraint of a tightly coupled connection defined in \eqref{14}. Therefore, a condition defined for edges ($\mathbb{E}=\{{\mathbb{E}}_{gj}|g\epsilon G,j\ \epsilon \widehat{{\mathfrak{S}}^q_g}\}$) will also fulfill all constraints mentioned in \eqref{14}, such that $j\epsilon \widehat{{\mathfrak{S}}_g}$ UAVs communicate with others following constraints defined in \eqref{14} and \eqref{18}, when $|\widehat{{\mathfrak{S}}_g|}=Q_g$, it supports a ring equivalent network which minimizes the maxima of the shortest path of ${G}^*$ as per edge limitation \eqref{14}. 

Mathematical expression proposed in corollary 1 requires subset relation between feasible and optimal network structures like ${G}^'\subseteq {G}^*$. Furthermore, edges of optimal network components comprise feasible network components with members denoted by $Q_g$. Proposed algorithm in Table II will explain edges connected inside optimal network topology. We denote $\mathbb{O}$ for algorithm complexity $\mathbb{O}(G^{q_g}$) at each airborne object $g$\textit{ }for the proposed optimal network. Therefore, mentioned complexity for the proposed topology is acceptable because each UAV operating in a distributed network is provided with very limited resources $\mathfrak{B}$. Consequently, our proposed network algorithm is purely distributed because each flying object shares its respective channel dataset of A2A link to the other 5G enabled UAVs. In this scheme, generally, the network is decentralized by excluding other extra 5G enabled UAVs from its own feasible network.

\subsection{Proposed Optimal learning Model using LSTM-DCGAN FOR 5G Enabled Maritime UAVs}
Following \cite{zhang2021distributed} and \cite{ferdowsi2019generative}, lemma 1 and proposition 1, for proposed optimal network topology $G^*$, distribution attained by the optimized generator ${\mathcal{G}}^*_g$  of each operating UAV, $g$ is denoted by \eqref{19}:
\begin{align} \label{19} 
& b^{{\mathcal{G}}^*}_g={\pi }_gb_g+\sum_{j\epsilon C_g}{{\pi }_{gj}log {(1-b^{\mathfrak{D}}_{j}b^{{\mathcal{G}}_{en}}_j})} \\
& b^{{\mathfrak{D}}^*}_g={\pi }_gb_g+log (\sum_{j\epsilon C_g}{1-{\pi }_{gj}log {(1-b^{{\mathfrak{G}}_{en}}_j}b^{{\mathcal{G}}_{en}}_j)}) \nonumber
\end{align}
 
Here it is clear that for each UAV $g$ having the angle of arrival (AoA) and angle of departure (AoD), the generator distribution function is equivalent to the combined effect of channel dataset distribution $b_g$with the generator distribution $b^{{\mathcal{G}}_{en}}_j$of objects from the set $C_j$. Therefore, in this situation discriminator will receive two datasets of the same channel with the same information i.e. one from a dataset from a local device and the second from the combined source from $H_g$ and $\{{{\mathcal{G}}_{en}}_{\left(j\right)}{\}}_{\forall j\epsilon C_g}.$ As result, the discriminator will not be able to differentiate between the real dataset and generated dataset, which makes the model equally likely for both 1 and 0 with a probability of 0.5. Therefore, we model discriminators output as in \eqref{20}:

\begin{equation} \label{20} 
{\mathfrak{D}}^*_g=\frac{b^{\mathrm{b}}_g}{b^{\mathfrak{D}}_g+f^{{\mathcal{G}}^*_{en}}_g}=\frac{1}{2} 
\end{equation} 

Now, we can say that convergence of UAV over local adversarial training towards $({{\mathcal{G}}_{en}}^*_g,{\mathfrak{D}}^*_g)$ is directly proportional to the convergence of dataset sharing within UAV network toward Nash Equilibrium \cite{ferdowsi2019generative}, given that distribution of each airborne object \textit{g }operating over mmWave link is learned i.e. ${{\mathcal{G}}_{en}}^*_g\sim b^*_g=b.$ Secondly, each UAV will only receive channel distribution \textit{b} from a generative model trained on the dataset from its optimal generator. Thus, we designed a data-driven method to verify optimal DCGAN distributed model $({{\mathcal{G}}_{en}}^*_g,\ {\mathfrak{D}}^*_g)$ elaborated in Table II: Proposed Algorithm.

We already mentioned that in LSTM-DCGAN, the network is overloaded with continuous collaboration $Load=\ I_{tr{(\mathcal{G}}_{en})}\sum_{g\epsilon G}{\eta H_g\rho Q_g=I_{tr{(\mathcal{G}}_{en})}\eta H\rho Q\mathfrak{B}}$, entailing all communication before aching convergence, such that the proposed optimal network has already minimized $I_{tr{(\mathcal{G}}_{en})}$. Hence, the learning mechanism proposed in Table II, is less complex with minimum communication overhead in optimal network architecture. 

Furthermore, for our LSTM-DCGAN, we improved transmission load by adjusting $\eta$ to fulfill other constraints in wireless communication. Similarly, algorithmic complexity big O of adversarial training for each UAVs operating locally is almost the same as the originally proposed in CGAN model in \cite{mirza2014conditional}. Whereas, proposed LSTM-DCGAN learning model has complexity nearly $I_{tr{\mathcal{G}}_{(en)}}$ times of originally proposed scheme. 

The other similar scheme was proposed in \cite{ferdowsi2019generative} about FL-CGAN, which introduced one central computing unit to calculate the average of values of generators and discriminators of each flying object within the network and reply accordingly with the updated value, on the contrary, our proposed model is fully distributed and decentralized with LSTM generator and discriminator. Similarly the \cite{zhang2021distributed} was a simple DCGAN approach with same complexity but it fails to handle synchronous learning at lower latency. The proposed work is able to provide decentralized method which effectively handles synchronous learning at much faster rate.  

\begin{table*}[!ht]\footnotesize
\renewcommand{\arraystretch}{1.1}
\centering
\caption{Proposed Algorithm}
\vspace*{-4pt}
\begin{tabular}{|p{7.0in}|} \hline 
\textbf{Proposed Algorithm: LSTM-DCGAN model-based learning for 5G Enabled Maritime UAV channel modeling} \\ \hline 
\textbf{Step1:}\begin{itemize} \item Each channel sample dataset $h_{gj}\ where\ j\epsilon G_{-g}$, collected by each airborne object \textit{g} and shares only a list of 5G enabled UAVs from a feasible set $\widehat{{\mathfrak{S}}_g}.$ \item Here we define a condition${\bigcup }^G_{g=1},\ \widehat{{\mathfrak{S}}_g}=G,\ such\ that\ |\widehat{{\mathfrak{S}}_g}|\ge Q_g$, if yes then move to the next step, otherwise, all UAVs will relocate them to position and perform previous operation again. \item On satisfying condition in above, define the network graph $G$ such that each edge is at $\left\{{\mathbb{E}}_{gj}\mathrel{\left|\vphantom{{\mathbb{E}}_{gj} g\epsilon G,j\epsilon \widehat{{\mathfrak{S}}_g}}\right.\kern-\nulldelimiterspace}g\epsilon G,j\epsilon \widehat{{\mathfrak{S}}_g}\right\}$ \end{itemize} \textbf{Step2:}\begin{itemize} \item Here we will check condition using the loop $\left|\widehat{{\mathfrak{S}}_g}\right|>Q_g$ for each $g$\textit{ }UAV,\textbf{ }Such that by removing edge ${\mathbb{E}}_{gj}$ from a set of edges $\mathbb{E}$ at a condition where $j={min}_{j\epsilon {\mathfrak{S}}_g}{\ell }^{max}_g\left(\mathbb{G}-{\mathbb{E}}_{gj}\right)-{\ell }^{max}_g\left(\mathbb{G}\right)$, ensuring (${\bigcup{}}_{\mathcal{I}\epsilon G_{-g}}\ \widehat{{\mathfrak{S}}_{\mathcal{I}})}\bigcup \left(\widehat{{\mathfrak{S}}_g}-j\right)=G$ and $\exists {\widehat{{\mathfrak{S}}_{\ }}}^q_g\subseteq \left(\widehat{{\mathfrak{S}}_{\mathcal{I}}}-j\right),\ \sum_{\mathcal{I}\epsilon {\widehat{{\mathfrak{S}}_{\ }}}^q_g}{P_{w\left(g\mathcal{I}\right)}\le P_{w\left(\mathrm{max}\right)};}$ \item Till loop iteration $\left|\widehat{{\mathfrak{S}}_g}\right|=Q_g$ for all \textit{g}$\epsilon G$\end{itemize} \textbf{Step3:} \begin{itemize} \item The first step in learning is to initialize each UAV's LSTM based discriminator and generator ${\mathfrak{D}}_g\ \mathrm{and\ }{\mathcal{G}}_{en(g)}$ \item Repeat: concurrently for all values of \textit{g}$\epsilon G$ \begin{itemize} \item Obtain sample \textit{o }for arrival and departure conditions ${\varphi }^1,{\varphi }^2,\dots ..{\varphi }^o\sim U\left[1,\ \mathcal{I}\right],\ \mathrm{and\ }u\ \mathrm{number\ of\ random\ inputs:\ }{\mathfrak{z}}^1,\dots ..{\mathfrak{z}}^o\sim b^{\mathfrak{z}}_g.$ \item  In the second step from the generator of each airborne object, we will generate a channel sample ${\mathcal{G}}_{en\left(g\right)}\left({\mathfrak{z}}^1\mathrel{\left|\vphantom{{\mathfrak{z}}^1 {\varphi }^1}\right.\kern-\nulldelimiterspace}{\varphi }^1\right),\dots \dots .{\mathcal{G}}_{en\left(g\right)}({\mathfrak{z}}^o|{\varphi }^o)$\item Sample $\pi ,\ o$ are real channel samples from the locally collected dataset: $\left\{h^1_g\mathrel{\left|\vphantom{h^1_g {\varphi }^1}\right.\kern-\nulldelimiterspace}{\varphi }^1\right\},\dots ..,\left\{s^{\pi ,o}_g\mathrel{\left|\vphantom{s^{\pi ,o}_g {\varphi }^{\pi ,o}_{\ }}\right.\kern-\nulldelimiterspace}{\varphi }^{\pi ,o}_{\ }\right\}\sim h_g$\item  Here we will update the LSTM parameter vector discriminator ${\emptyset }^d_g$ through the approximate conjugate gradient pursuit: ${\mathrm{\nabla }}_{{\emptyset }^{{\mathcal{G}}_{en}}_g}R\left({\mathfrak{D}}_g\left({\emptyset }^{\mathfrak{D}}_g\right)\right)=\frac{1}{2o}{\mathrm{\nabla }}_{{\emptyset }^{\mathfrak{D}}_g}[\sum^{{\pi }_g}_{\mathcal{I}=1}{{\mathrm{log} ({\mathfrak{D}}_g(s^{\mathcal{I}}_g|{\varphi }^{\mathcal{I}}+\sum^o_{\mathcal{I}=1}{\mathrm{log}\mathrm{}(1-{\mathfrak{D}}_g{\mathcal{G}}_{en\left(g\right)}\left({\mathfrak{z}}^{\mathcal{I}}\mathrel{\left|\vphantom{{\mathfrak{z}}^{\mathcal{I}} {\varphi }^{\mathcal{I}}}\right.\kern-\nulldelimiterspace}{\varphi }^{\mathcal{I}}\right)))+\sum_{j\epsilon C_i}{\sum^{{\pi }_{gj}o}_{\mathcal{I}=1}{{\mathrm{log} ({\mathfrak{D}}_g\left(h^{\mathcal{I}}_j\mathrel{\left|\vphantom{h^{\mathcal{I}}_j {\emptyset }^{\mathcal{I}}_{\ }}\right.\kern-\nulldelimiterspace}{\emptyset }^{\mathcal{I}}_{\ }\right))]\ }}}}\ }}$\item  Afterward, we will update LSTM parameters of generator via approximate conjugate gradient pursuit: ${\mathrm{\nabla }}_{{\emptyset }^{{\mathcal{G}}_{en}}_g}R\left({\mathcal{G}}_{en\left(g\right)}\left({\emptyset }^{{\mathcal{G}}_{en}}_g\right)\right)=\frac{1}{o}{\mathrm{\nabla }}_{{\emptyset }^{{\mathcal{G}}_{en}}_g}\sum^o_{\mathcal{I}=1}{{\mathrm{log} \left(1-{\mathfrak{D}}_g\left({\mathcal{G}}_{en}\left({\mathfrak{z}}^{\mathcal{I}}\mathrel{\left|\vphantom{{\mathfrak{z}}^{\mathcal{I}} {\emptyset }^{\mathcal{I}}}\right.\kern-\nulldelimiterspace}{\emptyset }^{\mathcal{I}}\right)\right)\right)\ }}$\end{itemize}\end{itemize}
Repeat this process till convergence to \textbf{Nash Equilibrium}. 
\newline  \\ \hline 
\end{tabular}
\vspace*{-4pt}
\end{table*}

Unlike our scheme, the FL-GAN model is dependent on the centralized controller. Furthermore, with available resources using the bulk amount of data is restricted in the FL-GAN model because transmission overhead of FL-GAN increase with the rise in adopted model size. On the other side, in our LSTM-Distributed CGAN scheme each neural network is trained by each respective UAV, hence, every model within the network can be completely different from others. Consequently, it is worth noting that results presented by \cite{ferdowsi2019generative} show distributed GAN outpaces both schemes FL-GAN and MD-GAN in the context of efficient communication and accuracy.

\begin{figure}[!t]
\centering
\includegraphics [width=0.50\textwidth]{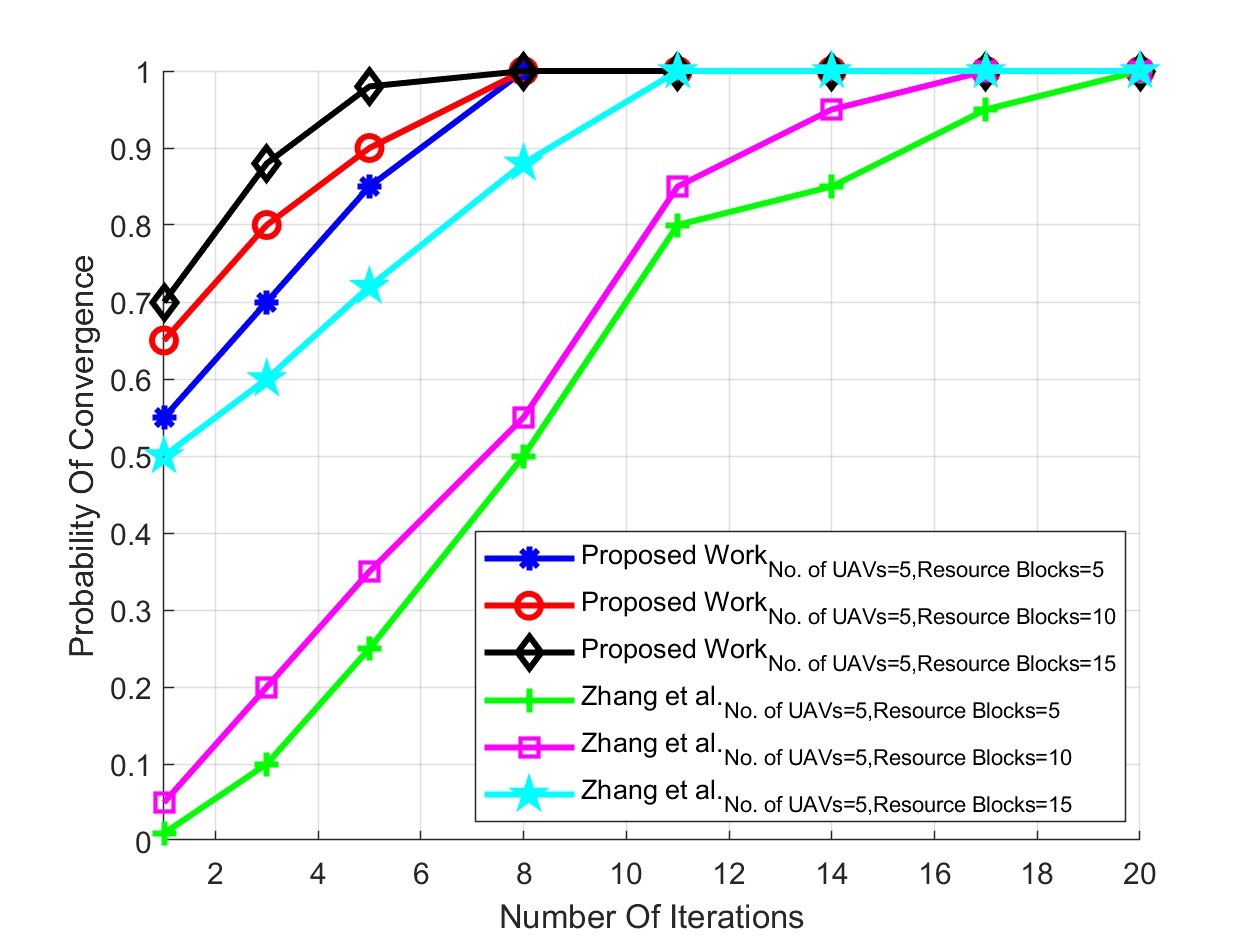}
\caption{Probability of Convergence Vs Number of Iterations (Changing Number of Resource Blocks).}
\label{fig:Fig3}
\end{figure}

\section{Performance Evaluation of The Proposed Work}

We simulate our proposed network with a number of maritime UAVs \textit{G=}5 with allocated resource blocks $\mathfrak{B}=5$ to facilitate a 5G enabled maritime network. We directed that each airborne object only collects data set from the region which is non-overlapping with the samples collected by other 5G enabled maritime UAVs in the set \textit{G}. Other parameters used in the simulation are with transmitter array \textit{M= }256 with receiver array have \textit{N=}64, operating in $\mathcal{I}=81$directions, at frequency 30 GHz and ${\omega }_b=2$MHz with maximum power $P_{w_{max}}=40dBm $ at noise power ${\sigma }^2=-174dBm/Hz$ and assumed training error $T_{error}=0.01$ with LSTM-DCGAN probability $P_{rob\_{LSTM-DCGAN}}=0.99$ with SNR threshold $T_{Th}=12 dB$, $t_{DCGAN}=$$0.01$ seconds convergence time with constants $\eta =0.5,\ \rho =11$ and depth of dataset from each 5G enabled UAV$_g$ is $H_g=10000$. Extensive comparsion between the proposed work and previous state of the art works has been done to evaulte the perfromance of the proposed work. 

\begin{figure}[!t]
\centering
\includegraphics [width=0.50\textwidth]{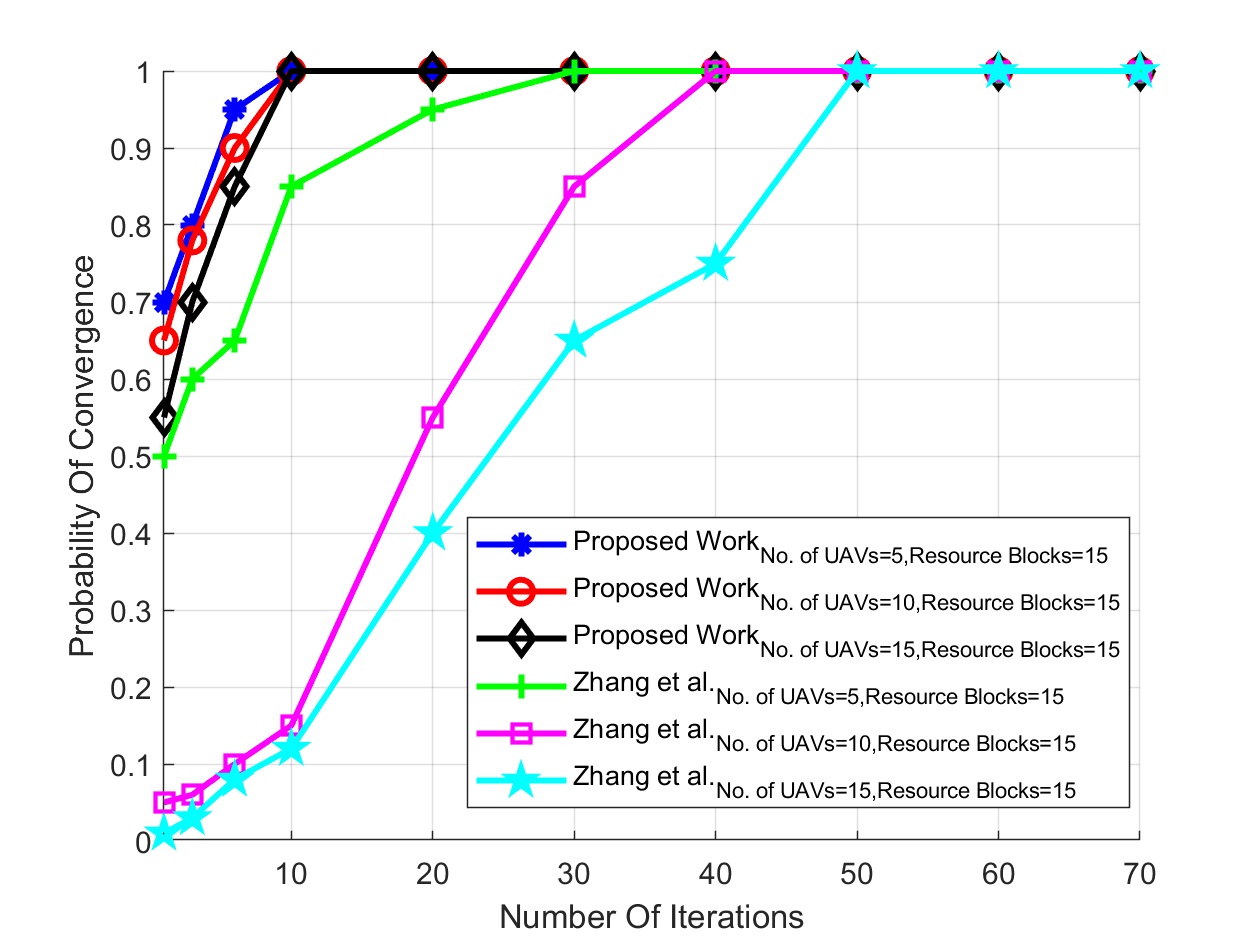}
\caption{Probability of Convergence Vs Number of Iterations (Changing Number of 5G Enabled UAVs).}
\label{fig:Fig4}
\end{figure}

\subsection{Convergence Anaylsis of the Proposed Work}

Fig.3 shows simulation results for Convergence probablility with respect to the number of iterations with changing number of resource blocks. We have used convergence of NE as KPI to assess the learning rate of our proposed model. In this case the fixed value of 5G enabled UAVs i.e.  $G=5$ with the variable number of resources varying $\mathfrak{B}= $ 5 to 15, result shows that convergence rate also increases with the rise in the number of allocated resources. From the Fig. 3 it can be clearly seen that proposed work achieved higher convergence probablility with respect to the previous state of art method \cite{zhang2021distributed}. This higher convergnece would mean that proposed work would be much suited for mission crtical applications which uses 5G enabled UAVs. 

In Fig.4, we have kept the number of resources as constant i.e. 15 with variable UAVs to observe the convergence of the proposed model. The number of UAVs being used by network are 5,10,15 respectively and the result shows that with the increase in the number of airborne objects the network becomes more and more complex with increased path length among UAVs. Therefore, the overall convergence rate is inversely proportional to the number of UAVs in-network, because learning in early iterations contains large and inefficient data sharing which restricts fast convergence. 

\subsection{Learning Performance of the Proposed Work}
In above part we evaluated our model in terms of convergence time. In this subsection we will focus on evaulating the learning performance of LSTM-DCGAN (proposed work) in comparison with five available models i.e. independent CGAN at each UAV without cooperation \cite{xia2020generative}, CGAN model with central controller collecting raw data from all nodes \cite{cheng2020modeling}, FL-CGAN \cite{elbir2021federated}, MD-CGAN distributed learning scheme \cite{ye2020deep}, and Zhang et al. \cite{zhang2021distributed}. We have used the Average Jensen-Shannon (JSD) as a performance indicator for learning accuracy; such that for higher accuracy, JSD must have a minimum value. From the Fig.5 we can clearly show that the proposed work achieved the minimum average JSD value as compared to the previous state of the art works. Moreover, our proposed method is independent of the central controller, which leads to having a more robust network, achieve higher accuracy, and vigorous toward failure when compared with MD-CGAN and FL-CGAN. 

\begin{figure}[!t]
\centering
\includegraphics [width=0.50\textwidth]{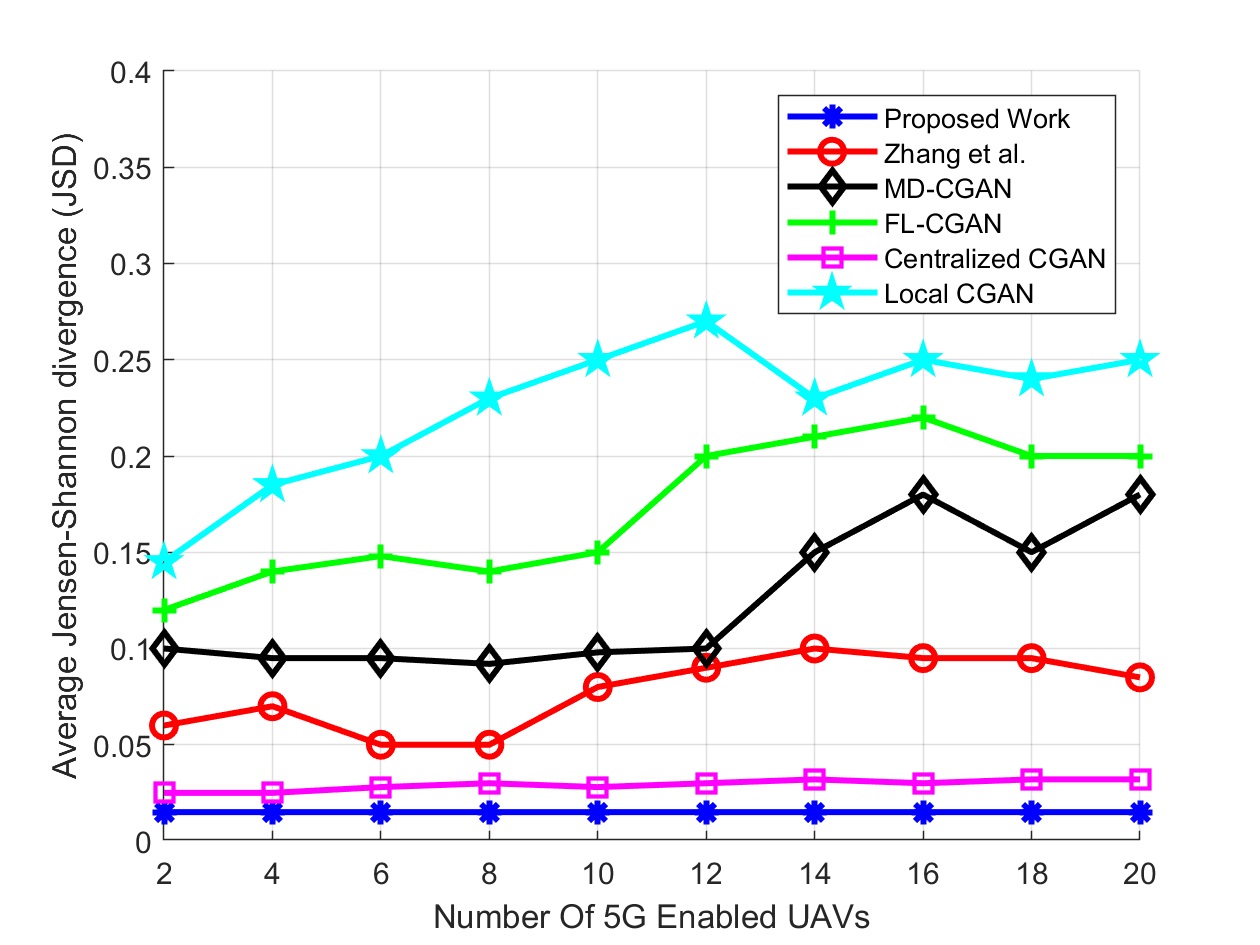}
\caption{Communication Overhead Vs Number of Resource Blocks.}
\label{fig:Fig5}
\end{figure}

\subsection{Communication Overhead of the Proposed Work}
One of the major problems for previous resource allocation methods was that as the Number of resources blocks increases the communication overhead also increases. This means that the proposed method should overcome this drawback as for 5G enabled UAV communication lower latency is mandatory. From Fig.6 it can be seen that the communication overhead of the proposed work almost remains the same as we are increasing the number of resource blocks. Moreover, communication overhead attained by the proposed method is minimum as compared to the existing state of artworks. 

\begin{figure}[!t]
\centering
\includegraphics [width=0.50\textwidth]{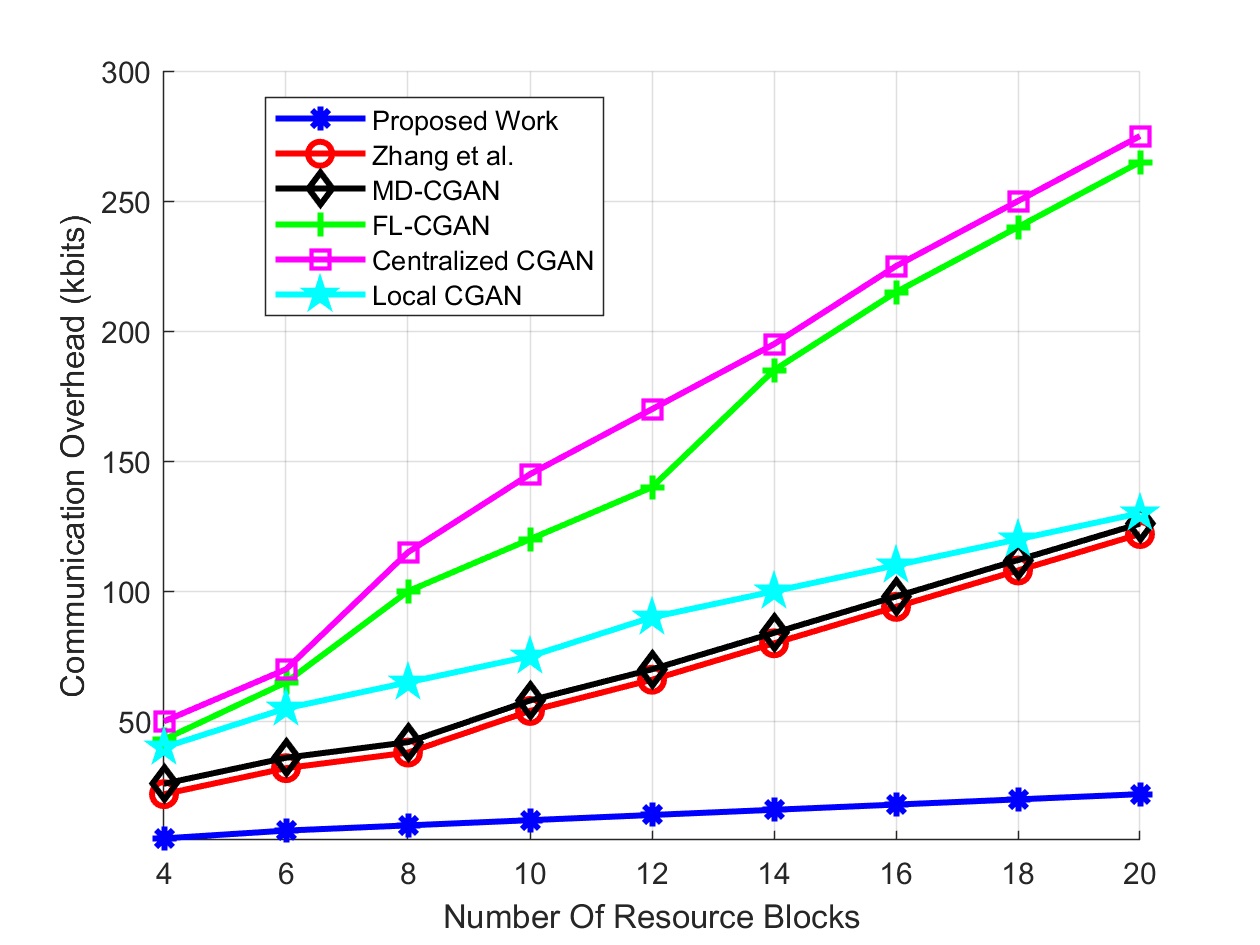}
\caption{Communication Overhead Vs Number of Resource Blocks.}
\label{fig:Fig6}
\end{figure}

\begin{figure}[!t]
\centering
\includegraphics [width=0.50\textwidth]{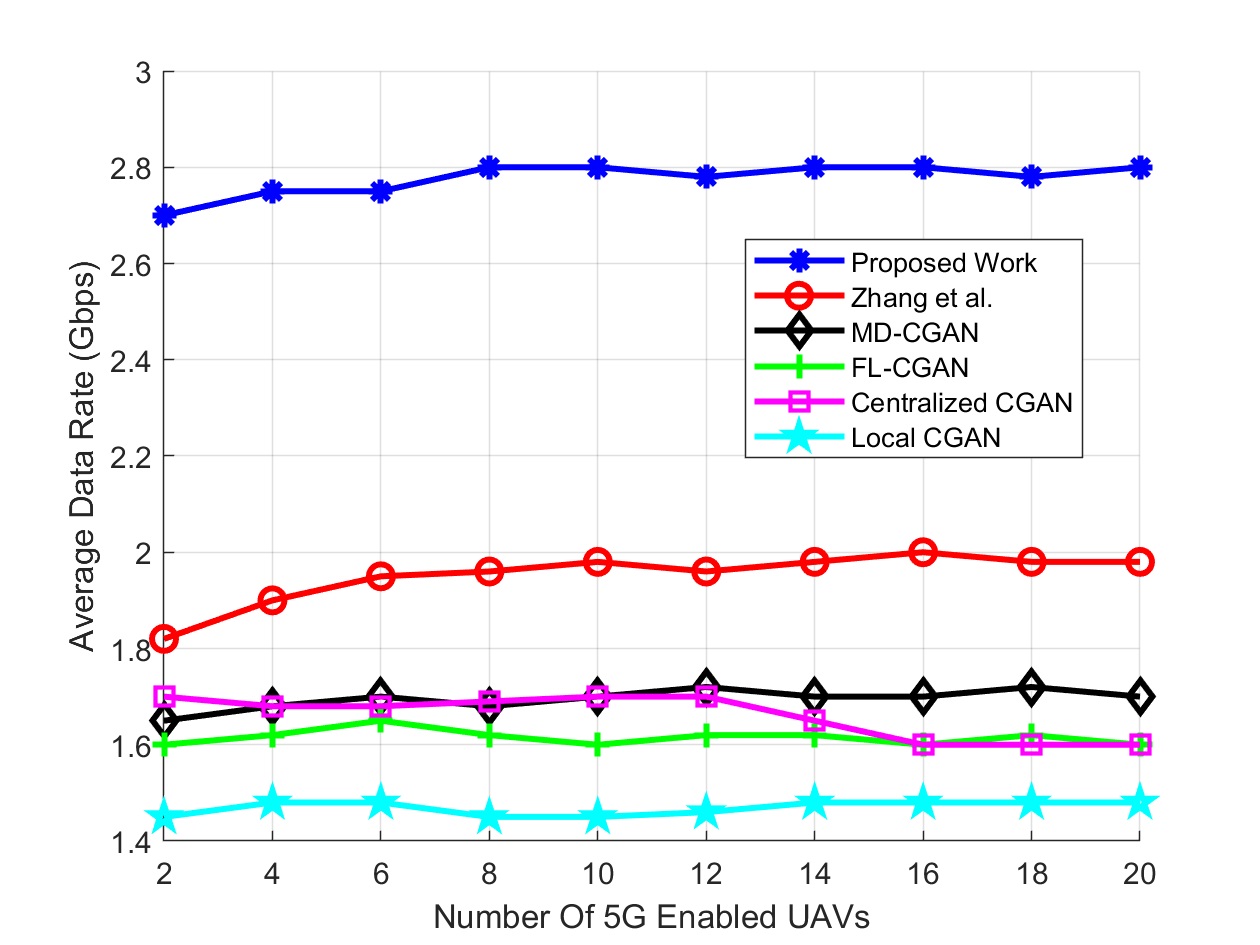}
\caption{Average Data Rate Achieved.}
\label{fig:Fig7}
\end{figure}
\subsection{Data Rate of the Proposed Work}
Fig.6 shows the average data rate achieved with respect to the 5G enabled UAVs. Form the Fig.7 it can be clearly seen that the proposed work outperforms the previous works in attaining the higher data rate. More importantly average data rate remains high and constant as the number of 5G enabled UAVs increases.

\section{Conclusion}
In this work we present an estimation of the channel model for 5G enabled maritime UAVs network. There are three key contributions initially, we formulated channel estimation method which directly aims to adopt channel state information (CSI) of mmWave from the channel model inculcated by UAV operating within the Long Short Term Memory (LSTM)-Distributed Conditional generative adversarial network (DCGAN) i.e. (LSTM-DCGAN) for each beamforming direction. Secondly, to enhance applications for the proposed trained channel model for the spatial domain, we have designed an LSTM-CGAN based UAV network, where each one will learn mmWave CSI for all distributions. Lastly, we categorized the most favorable LSTM-DCGAN training method and emanated certain conditions for our UAV network to increase the channel model learning rate. Simulation results have shown that at each UAV, our proposed LSTM-DCGAN based network is vigorous to the error generated through local training. Lastly the proposed work has been compared with the other available state-of-the-art CGAN network architectures i.e. stand-alone CGAN (without CSI sharing), Simple CGAN (with CSI sharing), multi-discriminator CGAN, and federated learning CGAN. Simulation results have shown that the proposed LSTM-DCGAN structure demonstrates higher accuracy during the learning process and attained more data rate for downlink transmission as compared to the previous state of artworks.


\ifCLASSOPTIONcaptionsoff
  \newpage
\fi

{\small
\bibliographystyle{IEEEtran}
\bibliography{references_Paper}
}

\end{document}